\documentclass[pdflatex,sn-mathphys-num]{sn-jnl}

\usepackage{graphicx}%
\usepackage{multirow}%
\usepackage{amsmath,amssymb,amsfonts}%
\usepackage{amsthm}%
\usepackage{mathrsfs}%
\usepackage[title]{appendix}%
\usepackage[table]{xcolor}%
\usepackage{textcomp}%
\usepackage{manyfoot}%
\usepackage{booktabs}%
\usepackage{algorithm}%
\usepackage{algorithmicx}%
\usepackage{algpseudocode}%
\usepackage{listings}%
\usepackage{array}

\theoremstyle{thmstyleone}%
\theoremstyle{thmstyletwo}%
\newtheorem{example}{Example}%

\theoremstyle{thmstylethree}%
\newtheorem{definition}{Definition}%

\newcommand{\twod}{\ensuremath{\{0, 1\}}}

\newcommand{\any}[0]{{\star}}

\newcommand{\node}[1]{\texttt{#1}}

\usepackage{xspace}

\newcommand{\ldoi}[0]{\text{LDOI}\xspace}

\newcolumntype{L}[1]{>{\raggedright\let\newline\\\arraybackslash\hspace{0pt}}m{#1}}
\newcolumntype{C}[1]{>{\centering\let\newline\\\arraybackslash\hspace{0pt}}m{#1}}
\newcolumntype{R}[1]{>{\raggedleft\let\newline\\\arraybackslash\hspace{0pt}}m{#1}}

\begin{document}

\title[Relating biomarkers and phenotypes using dynamical trap spaces]{Relating biomarkers and phenotypes using dynamical trap spaces}









\author[1]{\fnm{Samuel} \sur{Pastva}}\email{xpastva@fi.muni.cz}

\author[2]{\fnm{Kyu Hyong} \sur{Park}}\email{kjp5774@psu.edu}

\author[3]{\fnm{Jordan C.} \sur{Rozum}}\email{jordan.rozum@pnnl.gov}

\author[4,*]{\fnm{Van-Giang} \sur{Trinh}}\email{van-giang.trinh@hcmut.edu.vn}

\author[2,*]{\fnm{R\'eka} \sur{Albert}}\email{rza1@psu.edu}

\affil[1]{\orgdiv{Faculty of Informatics}, \orgname{Masaryk University}, \orgaddress{\street{Botanicka 68a}, \city{Brno},~\postcode{60200}, \country{Czech Republic}}}

\affil[2]{\orgdiv{Department of Physics}, \orgname{Pennsylvania State University}, \orgaddress{\street{251 Pollock Road}, \city{University Park}, \postcode{16803}, \state{Pennsylvania}, \country{USA}}}

\affil[3]{\orgdiv{Biological Sciences Division}, \orgname{Pacific Northwest National Laboratory}, \orgaddress{\street{902 Battelle Blvd}, \city{Richland}, \postcode{99352}, \state{WA}, \country{USA}}}

\affil[4]{\orgdiv{Faculty of Computer Science and Engineering}, \orgname{Ho Chi Minh City University of Technology (HCMUT), VNU-HCM}, \city{Ho Chi Minh City}, \country{Vietnam}}

\abstract{Connecting the dynamics of biomolecular networks to experimentally measurable cell phenotypes remains a central challenge in systems biology. Here we introduce a model-based definition of phenotype as a partial steady state that is committed to a certain dynamical outcome while otherwise being minimally constrained. We focus on Boolean models and define \emph{dynamical phenotypes} as complete trap spaces that maximally specify a chosen set of phenotype-determining nodes that correspond to biomarkers while keeping external inputs unconstrained. We show that dynamical phenotypes can be efficiently identified without full attractor enumeration. Using four published models, including a 70-node Boolean model of T cell differentiation, we show that dynamical phenotypes recover known cell types and activation states, and indicate the environmental conditions ensuring their existence. We also propose a method to identify informative phenotype-determining nodes based on the canalization of the Boolean functions. This method reveals biologically relevant cell state information that is complementary to the phenotypes manually defined by model creators and is validated by two attractor-based approaches. Our results demonstrate that dynamical phenotypes provide a scalable framework for linking model structure, external inputs, and phenotypic outcomes, and offer a principled tool for model-guided biomarker selection.}




\maketitle

\section{Introduction}
The characterization of a biologically observable cell phenotype typically involves the values of a relatively small number of readouts (\emph{biomarkers}). Phenotype occurrence is generally explained by the cell's response to one or a few known external inputs. Considering only the values of a small number of biomarkers may lead to an insufficient specification, missing important distinctions among cell states that have the same observed marker values. At the other end of the spectrum of characterization, high-throughput transcriptome, proteome, and metabolome assays indicate comprehensive information about the cell state. Yet, it is challenging to pinpoint what fraction of this information is relevant to the phenotype in question. Here, we propose a model-informed method to identify biologically meaningful intermediate specifications of a cell phenotype.

The steps taken toward a more focused explanation of a cell phenotype often involve the concept of a \emph{pathway}, which is a set of regulatory relationships between biomolecules represented by a directed, signed graph. Signaling pathways included in repositories such as KEGG~\cite{kanehisa2025kegg} or Reactome~\cite{fabregat2018reactome} typically use extracellular molecules as inputs and have output nodes that represent abstract proxies of a phenotype (e.g., an ``Apoptosis'' or ``Proliferation'' node). Such a pathway may be used to explain the occurrence of a cell phenotype under an external input combination by observing reachability between the input(s) and the phenotype node(s).

Pathways that govern different phenotypes are thought to constitute modules of a genome-wide biomolecular network~\cite{hartwell1999modular}. Despite sustained efforts to identify modules using graph-theoretic approaches, the emerging conclusion is that functional modules are not detectable from the network structure alone~\cite{alexander2009understanding}. Only via a description of the dynamics of information propagation, embodied in a mathematical model, can one identify functional modules~\cite{alexander2009understanding, kadelka2020modularity}. 

A common modeling practice is to focus on a pathway or module that determines one or a few phenotypes \cite{tyson2022timekeeping}. Each node in the pathway is characterized by a state, which changes due to the interactions within the system. Dynamic models use continuous or discrete state variables and may incorporate stochasticity. Each modeling framework has its benefits and most appropriate use cases \cite{wooten2017mathematical,ma2024comprehensive}. Here we focus on Boolean models, which are a popular and highly scalable type of qualitative dynamical model that can provide testable predictions about event ordering and perturbation response~\cite{abou-jaoude2016logical,rozum2024boolean}.

Dynamical models can be used to explain phenotypes in much the same way as experimental approaches: by relating input conditions to response outputs~\cite{helikar2008emergent,mai2009apoptosis,lu2015network, eduati2020patient}. Modelers select output nodes based on biological relevance and potential experimental observability, often with the implicit assumption that these nodes trigger downstream processes responsible for the physical manifestation of the phenotype. For example, a 41-node Boolean model of apoptosis versus cell survival incorporates a generic growth factor and TNF as external inputs~\cite{mai2009apoptosis}, and considers the sustained activation of the abstract node ``DNADamageEvent'' as a marker of apoptosis.  A subsequent continuous logic model of the same process~\cite{eduati2020patient}, which also incorporates patient-specific data, instead uses the level of the caspase Cas3 as a readout of the apoptosis propensity.  A 70-node Boolean model of colitis-associated colon cancer~\cite{lu2015network} has two abstract phenotype nodes, ``Apoptosis'' and ``Proliferation'', and additionally designates STAT3, NFKB and $\beta$-catenin as phenotype marker nodes. 

Most such models have been analyzed primarily through simulations, with results communicated in terms of average values of phenotype-marking output nodes. While comparisons of these average values across different input conditions or under perturbations yield interpretable results, focusing exclusively on selected outputs obscures important aspects of the underlying dynamics. The most complete characterization of the long-term behavior of a dynamical system is given by its repertoire of attractors. By definition, an attractor specifies the value (or recurring temporal pattern) of all the nodes in the model, including the inputs. However, states that differ only in the values of external inputs are equivalent internal states. In many cases, external inputs are sensed by specific receptor proteins. States that only differ in the values of external inputs and their respective receptors may likewise be considered dynamically equivalent. These observations suggest that the most useful characterization of a phenotype should lie between the extremes of specifying only a small set of biomarkers and fully specifying the values of all model components.

Previous work addressed the topic of the most appropriate intermediate phenotype specification to some extent ~\cite{cohen2015mathematical,puniya2018mechanistic,wooten2021mathematical}.  For a few dynamic models in the literature, the authors identified the attractors of the whole system and then organized them into groups based on the values of designated phenotype-determining nodes (PDNs). Up to a handful of nodes were typically chosen by the modelers as markers or determinants of the phenotype. For example, a 38-node Boolean model of T cell differentiation in a variable environment~\cite{puniya2018mechanistic} defines four lineage-specific transcription factors as phenotype markers. The authors identified by simulations the expression patterns of these transcription factors and the number of ergodic sets in the state space that yield each of these patterns.  Another example is a 32-node Boolean model of the molecular pathways enabling tumor cell invasion and migration~\cite{cohen2015mathematical}, which identifies nine attractors by simulations and groups them into four phenotypes based on the values of 6 abstract phenotype nodes such as ``Migration''. Identifying the attractor repertoire of the system is a prerequisite to this type of phenotype identification. This can be computationally burdensome because biological Boolean models can have hundreds or even thousands of attractors per input combination~\cite{trinh2025mapping}.

Here we propose that a phenotype corresponds to a minimally constrained cell state that is committed to a specific configuration of biomarkers. We introduce a dynamical representation of a phenotype as the mathematical object known as a complete trap space. Informally, a complete trap space is a partial steady state in which the effects of the fixed node values have been fully propagated.  In particular, a dynamical phenotype corresponds to a complete trap space in which the values of inputs are maximally free to vary and the values of designated PDNs are as specified as possible. To identify these spaces without enumerating all attractors of the system, we introduce a symbolic method based on binary decision diagrams. This method does not require full knowledge of network attractors and scales with the number of unique stable state combinations of PDNs. We construct the mapping between inputs and dynamical phenotypes by determining all the minimal input combinations that enable each phenotype and aggregating these into a Boolean function. 

Because of the efficiency of our symbolic methodology, it is feasible to compare the dynamical phenotypes that correspond to multiple alternative choices of phenotype-determining nodes. This comparative analysis allows the identification of PDN sets that are optimal within the constraints of experimental assays. We analyze two alternative situations for dynamical phenotype determination. The first is when prior biological knowledge can be used to designate a set of PDNs. The second is when, in the absence of such knowledge, the regulation incorporated in the Boolean functions of the model is used to identify PDNs. 

We begin by recalling important concepts in Boolean modeling that lay the foundation for the introduction of dynamical phenotypes, which we define in section~\ref{sec:dynamical-phenotype}. We next present the case studies from the systems modeling literature to which we apply our methods. In sections~\ref{sec:constructing-from-pdn} through \ref{sec:validation-of-pdns}, we perform a detailed comparative analysis of methods to generate PDNs and the consequent dynamical phenotypes. We demonstrate our methods by performing a comprehensive study of a model of helper T  cell differentiation~\cite{naldi2010diversity}. The original analysis of this model focused on a small fraction of input combinations, leading to an incomplete characterization of the attractor repertoire. We determine the dynamical phenotypes and complete input-phenotype map corresponding to a PDN set motivated by the original article. We identify a smaller PDN set from the Boolean functions of the model, and validate it and the corresponding dynamical phenotypes by comparing them to the results of two methods that use the full attractor repertoire of the model. We conclude that dynamical phenotype identification in models can be used as a tool to aid phenotyping in the lab.

\section{Background and key concepts}
\label{sec:background}

\begin{figure}
    \centering
    \begin{minipage}{0.12\linewidth}
        \centering
        \includegraphics[width=1.0\linewidth]{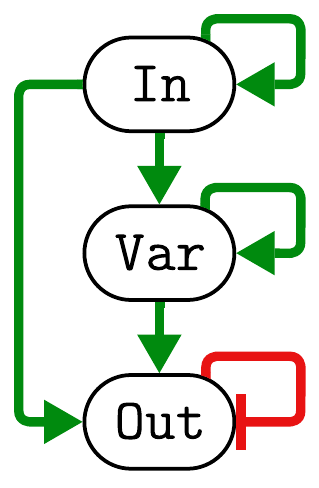}
    \end{minipage}
    \hspace{5pt}
    \begin{minipage}{0.4\linewidth}
        \centering
        \includegraphics[width=1.0\linewidth]{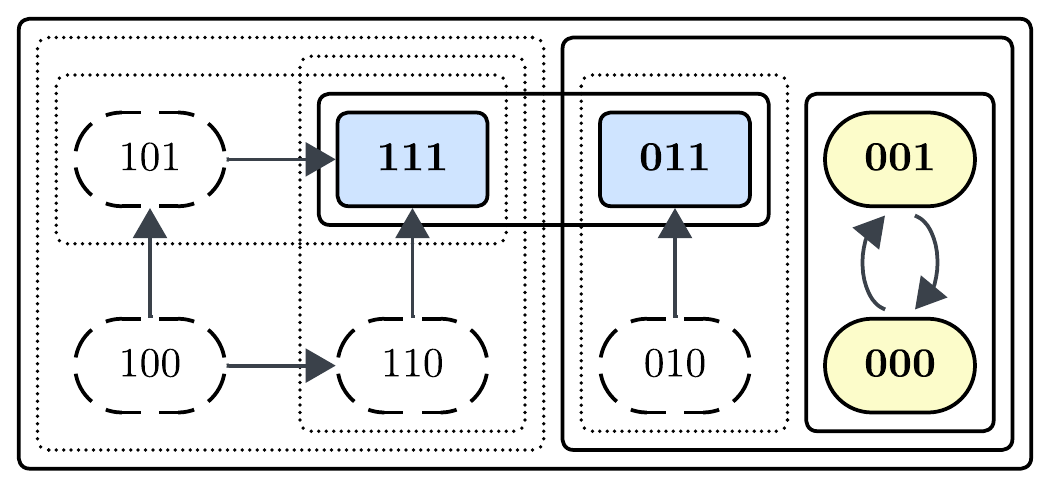}
    \end{minipage}
    \hspace{5pt}
    \begin{minipage}{0.22\linewidth}
        \centering
        \includegraphics[width=1.0\linewidth]{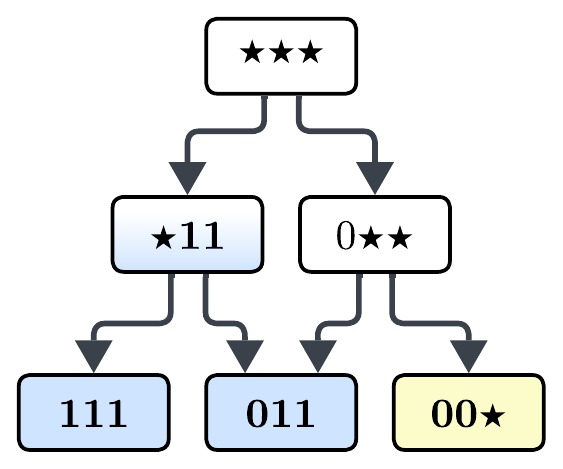}
    \end{minipage}

    \medskip
    \begin{minipage}{0.12\linewidth}
        \centering
        \small
        (a)
    \end{minipage}
    \hspace{5pt}
    \begin{minipage}{0.4\linewidth}
        \centering
        \small
        (b)
    \end{minipage}
    \hspace{5pt}
    \begin{minipage}{0.22\linewidth}
        \centering
        \small
        (c)
    \end{minipage}
    \medskip 
    
    \caption{Characterization of the simple Boolean model of Example~\ref{example:toy-network}. (a) Influence graph showing positive influences as green edges that end in triangle arrows, and negative influences as red edges that end in flat arrows. (b) Asynchronous state transition graph, showing two fixed points ($011$ and $111$) and a complex attractor ($100 \leftrightarrow 101$). The rectangular outlines indicate the model's repertoire of trap spaces, with continuous outlines corresponding to complete trap spaces. (c) Succession diagram consisting of all complete trap spaces; the directed edges depict subspace inclusion. The non-minimal complete trap space $\any11$, wherein the network is committed to $\node{Out}=1$, is highlighted using a blue gradient.}
    \label{fig:toy-example}
\end{figure}

Before presenting the main results of the paper, let us introduce the key theoretical concepts using a simple running example. 

\subsection{Foundations of Boolean models}

Boolean modeling of biological systems originated with the pioneering work of S. Kauffman and R. Thomas~\cite{kauffman1969metabolic,thomas1973boolean}. Hundreds of Boolean models have been developed in the last decades and have been collected in various repositories~\cite{cell-collective,bbm}. Recent reviews of Boolean modeling describe the construction and analysis of such models and exemplify the insights gained from them~\cite{rozum2024boolean,abou-jaoude2016logical}. Boolean modeling of biological systems is part of a broader research field that studies the dynamical properties of Boolean networks. The term ``Boolean network'' refers to any Boolean dynamical system with an underlying network structure. Since our focus is on Boolean models of biological systems, we will use the term ``Boolean model''.

A \emph{Boolean model}~\cite{kauffman1969metabolic,rozum2024boolean} consists of $n$ \emph{nodes} (also called \emph{variables}), each characterized by a state taken from the Boolean domain \twod.
The nodes typically have names, but we also expect them to be ordered and thus can reference them using indices $[1, n]$.
Every Boolean model induces a \emph{state-space} of $2^n$ Boolean vectors corresponding to $\twod^n$, each of which may be represented as a string of 1s and 0s.
Given a system state $x \in \twod^n$, we write $x_{\node{Var}}$ (or $x_{i}$) to denote the value of node \node{Var} (or the $i$-th node) in $x$.

A \emph{subspace} $S$ in a Boolean model is a hypercube within $\twod^n$. That is, $S$ represents an assignment of values to a subset of nodes. Nodes that are not assigned a value in $S$ are called \emph{free} in $S$, and those that are assigned a value are called \emph{fixed}. We denote the assignment of a value $x_i$ to node $i$  by $i\mapsto x_i$. We express a subspace as a set of Boolean state vectors in $\twod^n$, and we write it as a string consisting of the symbols 1, 0, and $\any$ to denote nodes fixed to the value 1, fixed to the value 0, or free, respectively. For example, $10\any$ represents the one-dimensional hypercube $\{101, 100\}$ within $\twod^3$.

Each node $i \in [1,n]$ is characterized by an \emph{update function} denoted $f_i$. 
Each $f_i$ takes the system state as input, and gives the next state of the node as output (formally, $f_i: \twod^n \to \twod$).
The update functions are given using logical expressions evaluated over the states of the system. Boolean models of biological systems often employ nested canalizing functions~\cite{jarrah2007nested,kadelka2024meta}, which reflect the ability of biological systems to buffer fluctuations in genetic and environmental signals. In these functions, multiple regulators form a hierarchy such that strong (canalizing) regulators override the weaker regulators’ effect. We say a regulator of a node is (fully) canalizing if there is a value of the regulator for which the regulated node's update function becomes fixed, regardless of the values of its other regulators.

We say that node \node{A} \emph{influences} node \node{B} (written $\node{A} \to \node{B}$) if the update function $f_\node{B}$ has \node{A} as its \emph{essential} input. We say that an essential input $i$ positively regulates its target $j$ if $f_j\big |_{x_i=0}\leq f_j\big |_{x_i=1}$ regardless of any other influences on $j$. Similarly, $i$ negatively regulates $j$ if $f_j\big |_{x_i=0}\geq f_j\big |_{x_i=1}$. It is possible that $i$ neither positively nor negatively regulates $j$, in which case we say the influence is non-monotonic.
All influences collectively determine the \emph{influence graph} (also called \emph{interaction} or \emph{regulatory} graph) of the Boolean model. Node \node{A} is called \emph{input} if it is not influenced by any other node. It is usually assumed that an input maintains its state, and this is represented by the identity function. 

Finally, the model's dynamics are determined by an \emph{update scheme}~\cite{chatain2020concurrency}, which describes how to apply update functions if more than one node can be updated concurrently. We focus on the \emph{stochastic asynchronous} update, where exactly one node, selected uniformly randomly, is updated at a time. The directed graph with Boolean states as vertices and aforementioned asynchronous transitions as edges is called the \emph{state transition graph} of the Boolean model. 

\begin{example}
\label{example:toy-network}
Consider a simple Boolean model $B$ consisting of three nodes, which we refer to as \node{In}, \node{Var}, and \node{Out}. 
Then, let the update functions be $f_{\node{In}}(x) = x_\node{In}$, $f_{\node{Var}}(x) = x_\node{In} \lor x_\node{Var}$, and $f_{\node{Out}}(x) = x_\node{In} \lor x_\node{Var} \lor \neg x_\node{Out}$. Here, $\lor$ and $\neg$ denote the Boolean OR and NOT operators, respectively. The influence graph and the asynchronous state-transition graph of $B$ are shown in Fig.~\ref{fig:toy-example}~(a), and Fig.~\ref{fig:toy-example}~(b), respectively. 
\end{example}

\subsection{Long-term dynamics of Boolean models}

The long-term dynamics of Boolean models are often studied through \emph{trap sets} and \emph{trap spaces}. A trap set is a set of states that is invariant under model dynamics. Meanwhile, a trap space is a trap set that can be characterized by restricting a subset of the nodes to fixed values~\cite{klarner2015computing} (the remaining nodes remain \emph{free}). In other words, a trap space is a trap set that is also a subspace. Notably, while the trap sets of a Boolean model depend on the update scheme used, its trap spaces do not~\cite{zanudo2013effective, pauleve2020reconciling}. Each trap space corresponds to a minimal, self-reinforcing network of node states called a stable motif. Stable motifs can be identified, for example, using a hypergraph or Petri-net representation of the update functions~\cite{zanudo2013effective, rozum2021parity,trinh2025mapping}.
Inclusion minimal trap spaces are of particular interest, as they represent parts of the state space that cannot be escaped, but that also cannot be further constrained into a smaller trap space.

Modelers are typically primarily interested in the model's \emph{attractors}, i.e., inclusion-minimal trap sets. Single-state attractors are called \emph{fixed points} (or \emph{steady states}) and multi-state attractors are referred to as \emph{complex attractors}. 
Note that every \emph{inclusion-minimal} trap space contains at least one attractor. In very rare cases, attractors can also appear outside of minimal trap spaces; such motif-avoidant attractors have not been observed in Boolean models of biological systems~\cite{THB2022,pastva2025open}.

We define that a trap space is called \emph{complete} if for every free node \node{Var}, the trap space contains \emph{some} states $x$ and $y$ such that $f_\node{Var}(x) \not= f_\node{Var}(y)$ (i.e., $f_\node{Var}$ is not constant within the trap space). 
Complete trap spaces are invariant under \emph{value percolation} (successively ``plugging in'' fixed node values into update functions via two-valued logic). Complete trap spaces correspond to the vertices of the so-called \emph{succession diagram}, which, as defined in~\cite{trinh2025mapping}, is a directed acyclic graph whose edges indicate inclusion.

\begin{example}
\label{example:attractors-traps}
    The Boolean model from Example~\ref{example:toy-network} admits two fixed points ($011$ and $111$) and one complex attractor ($000 \leftrightarrow 001$). It also admits ten trap spaces, specifically $\any\any\any$, $0\any\any$, $1\any\any$, $1\any1$, $11\any$, $00\any$, $01\any$, $\any11$, $011$, and $111$. Here, $\any$ denotes a free node so that, e.g., $01\any=\{010,011\}$. Six of these ten trap spaces are complete. 
    Three complete trap spaces exactly match the attractors (specifically 011, 111, and $00\star$), and the remaining three are $\any11$, $0\any\any$, and $\any\any\any$. 
    All attractors and trap spaces are visualized in Fig.~\ref{fig:toy-example}~(b). 
    The succession diagram, which summarizes the inclusion relations of complete trap spaces, is shown in Fig.~\ref{fig:toy-example}~(c).
\end{example}

\section{Our proposed representation of biological phenotypes in Boolean models}
\label{sec:dynamical-phenotype}

In this paper, we examine the relationship between biological phenotypes and the long-term behavior of the corresponding Boolean models. 
We assume that a subset of nodes can serve as determinants or markers of the phenotype; we call them \emph{phenotype-determining nodes} (PDNs). 
In the literature, these are typically designated by model authors based on experimental observations~\cite{cohen2015mathematical,dahlhaus2016boolean,helikar2008emergent,lu2015network,mai2009apoptosis}. Once PDNs are known, a subsequent analysis step is to determine the values of these nodes across attractors (or minimal trap spaces) to enumerate all admissible phenotypes. However, we argue that additional information can be gained by more targeted analysis, especially if the number of attractors (or minimal trap spaces) is large, e.g., due to the presence of many network inputs.

Here, we propose that phenotypes should be represented as (a subset of) complete trap spaces. This naturally groups attractors and minimal trap spaces into an intermediate-scale representation and systematizes the practice~\cite{wooten2017mathematical,cohen2015mathematical} of manually grouping attractors according to the values of output nodes. This motivates our definition of a \emph{dynamical phenotype}:

\begin{definition}[dynamical phenotype]
A \emph{dynamical phenotype subspace} $X$ is a complete trap space such that (1) all trap spaces included in $X$ share the same values of phenotype-determining nodes, and (2) there is no complete trap space containing $X$ with property (1). A \emph{dynamical phenotype} is the union of all dynamical phenotype subspaces sharing the same PDN values.
\end{definition}

Intuitively, a dynamical phenotype subspace represents a minimally constrained cell state that is committed to a specific configuration of biomarkers. Meanwhile, a dynamical phenotype is the trap set that combines all such complete subspaces where the PDNs are fixed to a specific combination of values.

Our rationale for this definition of the dynamic phenotype is the following: 
First, by not requiring the trap set to be fully minimal, we naturally allow the grouping of multiple related attractors (and minimal trap spaces) into one dynamical phenotype. 
Second, we focus on complete trap spaces (instead of trap spaces in general), because we expect the canalizing processes leading to value percolation to always achieve an irreversible steady state. In other words, we expect that having a node state out of sync with the value of the node's update function will be rare in practice, and when it happens it will almost never be sustained. 
Finally, the aim behind point (2) of the definition is to find the minimal specificity (i.e., the least specific state possible for non-PDNs) that \emph{guarantees} certain observable behavior.

Every dynamical phenotype must contain at least one minimal trap space with matching PDN values, and conversely, every minimal trap space is contained within some dynamical phenotype with matching PDN values. As such, the number of dynamical phenotypes is at most the number of minimal trap spaces. However, depending on the choice of PDNs and the behavior of the Boolean model, we expect the number of dynamical phenotypes to be often much smaller in practice.

\begin{example}
\label{example:phenotype-trap-space}
    Consider the Boolean model from Example~\ref{example:toy-network} and its complete trap spaces (Fig.~\ref{fig:toy-example}). Let us assume \node{Out} is the sole phenotype-determining node of this network. Here, $\node{Out} = 1$ corresponds to the dynamical phenotype subspace $\any11$ while the oscillation $\node{Out} = \any$ corresponds to the dynamical phenotype subspace $00\any$. Notice how the attractors with the same value of \node{Out} ($011$ and $111$) naturally group into a larger phenotype trap space $\any11$. Furthermore, we can easily deduce from this result that the input node \node{In} does not influence the emergence of $\node{Out} = 1$, but is required to be off for $\node{Out} = \any$. Finally, since \node{Var} is fixed in $00\any$ but not due to value percolation from \node{In} (after fixing $\node{In} = 0$, $f_\node{Var}(x) = x_\node{In} \lor x_\node{Var}$ simplifies to $x_\node{Var}$), we see that the phenotype $\node{Out} = \any$ is not fully determined by the input node of the network. Instead, an internal stable motif (namely, the self-regulation of $\node{Var}$ that sustains $\node{Var} = 0$) has to lock in for $\node{Out} = \any$ to be guaranteed.
\end{example}

\section{Case studies}

We illustrate the concept of dynamical phenotype on four published models of biomolecular modules that determine cell fates. Three of these models~\cite{wooten2021mathematical, cohen2015mathematical, dahlhaus2016boolean} have a relatively small size (19 to 32 nodes) and few (2 to 4) inputs. The original analysis of these models identified all attractors and then grouped them. We compare the dynamical phenotypes with the attractor groups and identify the input combinations under which each dynamical phenotype exists. 

In addition to these smaller models, we analyze a discrete dynamic model by Naldi and collaborators ~\cite{naldi2010diversity}  with a larger size (65 nodes) and a large number of inputs (13), whose full attractor repertoire has not been reported before. This model describes the differentiation of Th0 cells into Th1, Th2, Th17 or Treg cells~\cite{naldi2010diversity}. The 65 nodes include 12 constant nodes that refer to cellular context, 13 input nodes that describe the presence of antigen-presenting cells and cytokines in the cellular microenvironment, and 5 nodes with three expression levels. Naldi and colleagues~\cite{naldi2010diversity} analyzed a 34-node reduced version of their model, which retains the 13 inputs. In simulations under a set of 8 prototypical input combinations, they observed 38 attractors and classified them into 17 groups based on biological knowledge. The expression of a specific master transcription factor was chosen as a marker of each cell type: Th1 cells express TBET, Th2 cells express GATA3, Treg cells express FOXP3 and Th17 cells express RORGT. The article also reported attractors in which two or three of the four transcription factors are expressed and interpreted them as hybrid cell types. For each cell type, expression of NFAT and the production of a cell-specific cytokine (IFNG, IL4, IL17, or TGFB) was used to describe the cell state as resting, activated, or anergic (exhausted). 

\begin{figure}[h!]
    \centering    
    \includegraphics[width=1.0\linewidth]{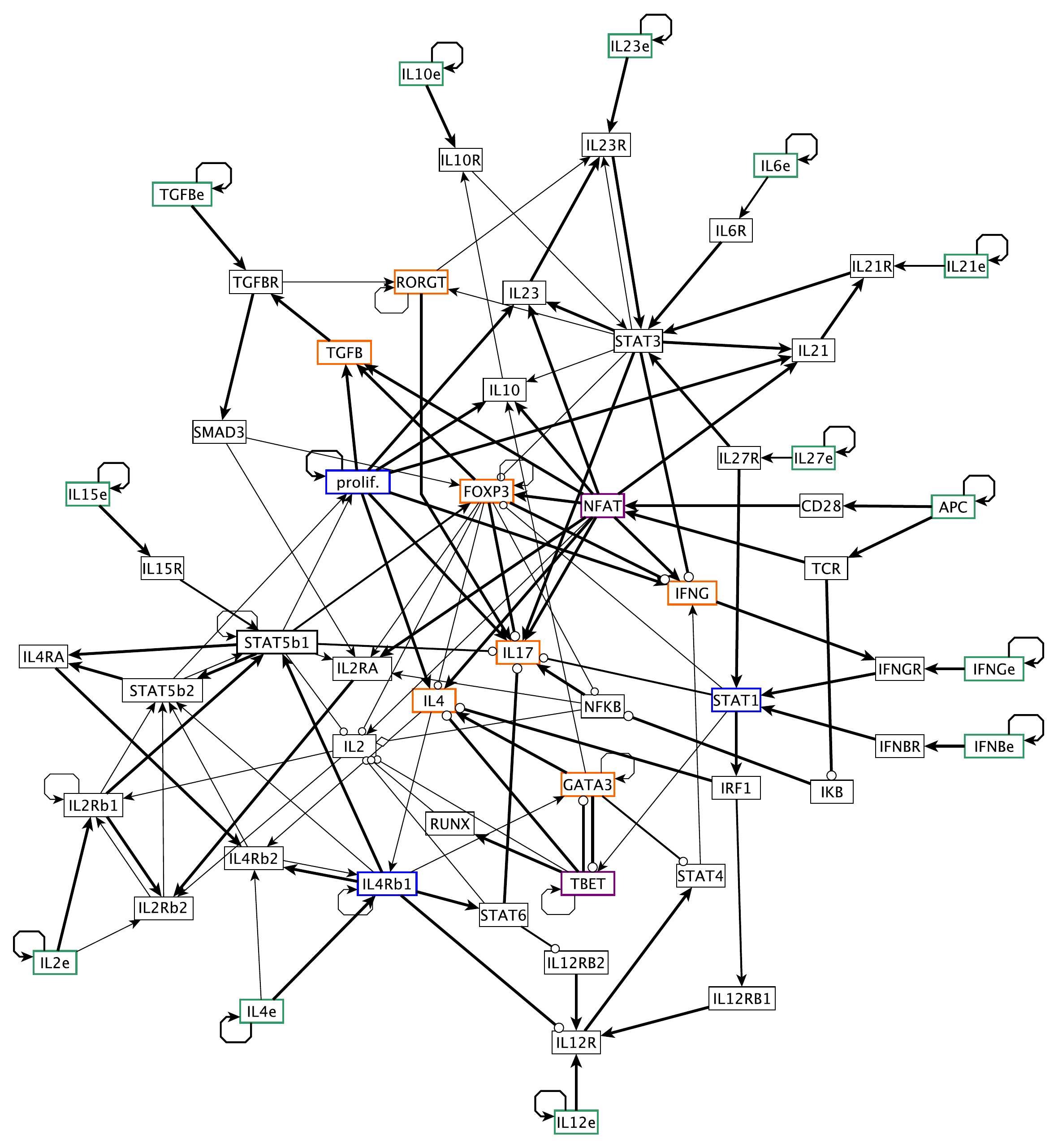}
    \caption{The influence graph of the 70-node Boolean version of the T cell differentiation model of Naldi et al. with 12 constant-value nodes removed.  Regular arrows represent positive influences (activation), and round arrows represent negative influences (inhibition). The thicker edges indicate canalizing regulators (either sufficient or necessary to elicit the activation of the target node). Red border indicates 7 biologically-motivated PDNs, blue border indicates 3 PDNs chosen based on regulatory logic, and purple border indicates 2 shared PDNs. Green borders indicate input nodes.}
    \label{Tcellnetwork}
\end{figure}

We analyze the 70-node Boolean version of the model available in the Biodivine Boolean Model (BBM) repository~\cite{bbm}. To create this model, the tool GINsim~\cite{ginsim} was used to represent each of the three-valued species (STAT5, IL2R,  IL4R, IL4RA, IL12RB1) with two Boolean nodes. This transformation is known to preserve attractors and complete trap spaces, hence it does not impact our result in comparison with the original paper. We then also substitute the reported value of the 12 constant nodes, which results in a 58-node Boolean model, whose influence graph is shown in Figure~\ref{Tcellnetwork}. Due to the $2^{13}$ input combinations, this model has a large number of attractors. At the time of the original study, it was not possible to identify the full attractor repertoire. Today, efficient tools exist to do so, such as \texttt{biobalm}~\cite{trinh2025mapping}. Thus, we will use this model for a comparative evaluation of complete trap space-based and attractor-based methods to identify phenotypes.

\section{Constructing dynamical phenotypes using known PDNs}
\label{sec:constructing-from-pdn}
We developed a symbolic method based on binary decision diagrams~\cite{Bryant86} to identify the complete trap spaces that make up dynamical phenotypes for a set of PDNs. This method's innovation is that it does not scale with the absolute number of complete trap spaces, but rather with the number of unique combinations of PDNs. We describe this procedure in Section \ref{phenotype_subspace_determination} (in the Methods).

We applied this procedure to three models reported in the literature whose original analysis included the identification of multiple attractors~\cite{cohen2015mathematical,wooten2021mathematical,dahlhaus2016boolean}. As described in detail in Appendix \ref{SI:three_examples}, we found that when using the phenotype-determining nodes chosen in each original article, the dynamical phenotypes agree with the attractor groups interpreted as phenotypes in the original article. The model of cancer cell invasion of Cohen and colleagues~\cite{cohen2015mathematical} has six abstract nodes: Metastasis, Apoptosis, CellCycleArrest, Migration, Invasion, and EMT. Using these nodes as PDNs, we identify four dynamical phenotypes, which agree with the four phenotypes identified in the original article: homeostasis, apoptosis, metastasis, and epithelial-to-mesenchymal transition (EMT). The model of yeast-to-hyphal transition of Wooten and colleagues~\cite{wooten2021mathematical} has three abstract nodes: hyphal\_initiation, hyphal\_maintenance, and HAG\_transcription. Specifying these nodes as PDNs, we recover the four phenotypes reported in the article: Yeast, Yeast-like, Hyphal-like, and Hyphal. The model of the Aurora Kinase in neuroblastoma by Dahlhaus and colleagues~\cite{dahlhaus2016boolean} has two dynamical phenotypes if designating the Aurora kinase as a PDN, in agreement with the asynchronous dynamics of the original model~\cite{park2023models}.

As our main case study, we determine the dynamical phenotypes of the 70-node Boolean version of the T helper cell differentiation model~\cite{naldi2010diversity} using the set of 9 phenotype-determining nodes that give rise to the phenotypes reported in the original analysis: TBET, GATA3, FOXP3, RORGT, NFAT, IFNG, IL4, IL17, and TGFB.

We identify 30 dynamical phenotypes across 145 dynamical phenotype subspaces (see also breakdown in Table~\ref{tab:phenotype-entropy-literature}). Each dynamical phenotype can be described as a CD4+ T cell type and activation state according to the classification used in the Naldi et al. article. Seventeen dynamical phenotypes correspond to the 17 attractor groups reported in the original paper. The remaining 13 dynamical phenotypes correspond to CD4+ T cell types not found in the original analysis, such as anergic Treg cells, resting Th17 cells, and Th1-Treg hybrid cells that express IFNG. Notably, we identify two dynamic phenotypes with oscillating TGFB expression. Some of these phenotypes (e.g. resting Th17 cells) were missed in the original analysis because the input combination necessary for them was not studied. Other phenotypes (e.g., anergic Treg cells) were missed because the original analysis sampled, rather than exhaustively explored, the state space. 

The 9 PDNs span a large range in terms of their representation in the 30 dynamical phenotypes: NFAT is expressed in 22 phenotypes, RORGT is expressed in 16, while IL17 is expressed in only one.
In Table~\ref{Bio_phenotypes} we list the 30 dynamic phenotypes by indicating the value of the 9 PDNs and the corresponding biological cell types/states.

\begin{table}[]
    \setlength{\tabcolsep}{3pt} 
    \centering
    \footnotesize
    \begin{tabular}{cccccccccl}
NFAT & TBET & IFNG & GATA3 & IL4 & FOXP3 & TGFB & RORGT & IL17 & biological phenotype\\\hline
0 & 0 & 0 & 0 & 0 & 0 & 0 & 0 & 0 & Th0\_re\\
0 &\cellcolor{green!25}1 & 0 & 0 & 0 & 0 & 0 & 0 & 0 & Th1\_re\\
0 & 0 & 0 & \cellcolor{green!25}1 & 0 & 0 & 0 & 0 & 0 & Th2\_re\\
0 & 0 & 0 & 0 & 0 & 0 & 0 & \cellcolor{green!25}1 & 0 & Th17\_re\\
0 & \cellcolor{green!25}1 & 0 & 0 & 0 & 0 & 0 & \cellcolor{green!25}1 & 0 & Th1-Th17\_re\\
0 & 0 & 0 & \cellcolor{green!25}1 & 0 & 0 & 0 & \cellcolor{green!25}1 & 0 & Th2-Th17\_re\\
\cellcolor{green!25}1 & 0 & 0 & 0 & 0 & 0 & 0 & 0 & 0 & Th0\_ac\\
\cellcolor{green!25}1 & \cellcolor{green!25}1 & \cellcolor{green!25}1 & 0 & 0 & 0 & 0 & 0 & 0 & Th1\_ac\\
\cellcolor{green!25}1 & \cellcolor{green!25}1 & 0 & 0 & 0 & 0 & 0 & 0 & 0 & Th1\_an\\
\cellcolor{green!25}1 & 0 & 0 & \cellcolor{green!25}1 & \cellcolor{green!25}1 & 0 & 0 & 0 & 0 & Th2\_ac\\
\cellcolor{green!25}1 & 0 & 0 & \cellcolor{green!25}1 & 0 & 0 & 0 & 0 & 0 & Th2\_an\\
\cellcolor{green!25}1 & 0 & 0 & 0 & 0 & \cellcolor{green!25}1 & \cellcolor{green!25}1 & 0 & 0 & Treg\_ac\\
\cellcolor{green!25}1 & 0 & 0 & 0 & 0 & \cellcolor{blue!25}Osc & \cellcolor{blue!25}Osc & 0 & 0 & Treg\_ac\_o\\
\cellcolor{green!25}1 & 0 & 0 & 0 & 0 & \cellcolor{green!25}1 & 0 & 0 & 0 & Treg\_an\\
\cellcolor{green!25}1 & 0 & 0 & 0 & 0 & 0 & 0 & \cellcolor{green!25}1 & 0 & Th17\_an\\
\cellcolor{green!25}1 & \cellcolor{green!25}1 & 0 & 0 & 0 & \cellcolor{green!25}1 & \cellcolor{green!25}1 & 0 & 0 & Th1-Treg\_ac\_TGFB\\
\cellcolor{green!25}1 & \cellcolor{green!25}1 & 0 & 0 & 0 & \cellcolor{green!25}1 & 0 & 0 & 0 & Th1-Treg\_an\\
\cellcolor{green!25}1 & \cellcolor{green!25}1 & \cellcolor{green!25}1 & 0 & 0 & 0 & 0 & \cellcolor{green!25}1 & 0 & Th1-Th17\_ac\_IFNG\\
\cellcolor{green!25}1 & \cellcolor{green!25}1 & 0 & 0 & 0 & 0 & 0 & \cellcolor{green!25}1 & \cellcolor{green!25}1 & Th1-Th17\_ac\_IL17\\
\cellcolor{green!25}1 & \cellcolor{green!25}1 & 0 & 0 & 0 & 0 & 0 & \cellcolor{green!25}1 & 0 & Th1-Th17\_an\\
\cellcolor{green!25}1 & 0 & 0 & \cellcolor{green!25}1 & 0 & \cellcolor{green!25}1 & 0 & 0 & 0 & Th2-Treg\_an\\
\cellcolor{green!25}1 & 0 & 0 & \cellcolor{green!25}1 & \cellcolor{green!25}1 & 0 & 0 & \cellcolor{green!25}1 & 0 & Th2-Th17\_ac\_IL4\\
\cellcolor{green!25}1 & 0 & 0 & \cellcolor{green!25}1 & 0 & 0 & 0 & \cellcolor{green!25}1 & 0 & Th2-Th17\_an\\
\cellcolor{green!25}1 & 0 & 0 & 0 & 0 & \cellcolor{green!25}1 & \cellcolor{green!25}1 & \cellcolor{green!25}1 & 0 & Treg-Th17\_ac\_TGFB\\
\cellcolor{green!25}1 & 0 & 0 & 0 & 0 & \cellcolor{blue!25}Osc & \cellcolor{blue!25}Osc & \cellcolor{green!25}1 & 0 & Treg-Th17\_ac\_TGFB\_o\\
\cellcolor{green!25}1 & 0 & 0 & 0 & 0 & \cellcolor{green!25}1 & 0 & \cellcolor{green!25}1 & 0 & Treg-Th17\_an\\
\cellcolor{green!25}1 & \cellcolor{green!25}1 & 0 & 0 & 0 & \cellcolor{green!25}1 & \cellcolor{green!25}1 & \cellcolor{green!25}1 & 0 & Th1-Treg-Th17\_ac\_TGFB\\
\cellcolor{green!25}1 & \cellcolor{green!25}1 & 0 & 0 & 0 & \cellcolor{green!25}1 & 0 & \cellcolor{green!25}1 & 0 & Th1-Treg-Th17\_an\\
\cellcolor{green!25}1 & 0 & 0 & \cellcolor{green!25}1 & 0 & \cellcolor{green!25}1 & \cellcolor{green!25}1 & \cellcolor{green!25}1 & 0 & Th2-Treg-Th17\_ac\_TGFB\\
\cellcolor{green!25}1 & 0 & 0 & \cellcolor{green!25}1 & 0 & \cellcolor{green!25}1 & 0 & \cellcolor{green!25}1 & 0 & Th2-Treg-Th17\_an\\
    \end{tabular}
    \vspace{6pt}
    
    \caption{The 30 dynamical phenotypes based on the 9 biologically-motivated PDNs. The biological phenotype consists of a cell type and an activation state. We order the PDNs by first indicating the activity of NFAT, which distinguishes between resting and active/anergic cells, then indicating the transcription factor - cytokine pair for each of the four cell types (e.g., TBET and IFNG correspond to Th1 cells). In the shorthand notation of the phenotypes, ``re'' corresponds to resting, ``ac'' corresponds to active, ``an'' corresponds to anergic, ``o'' corresponds to oscillating. }
    \label{Bio_phenotypes}
\end{table}


The repertoire of dynamical phenotypes of the Naldi et al. T cell differentiation model implies that for each of the corresponding T cell states there exist input combinations and initial conditions under which the cell state is reachable. The obvious follow-up question is which input combination allows each phenotype. We address this question next.

\section{Dynamical phenotypes allow the understanding of the input-phenotype relationships}
\label{sec:input-phenotype results}

The original analysis of the T cell differentiation model by Naldi et al. ~\cite{naldi2010diversity} considered only 8 of the $2^{13}$ input combinations. Here we perform a comprehensive analysis, whose methodological details are described in Section~\ref{phenotype_subspace_determination}. Briefly, for each dynamical phenotype, we determine the input constraints required to observe it and express them as a logical condition in disjunctive normal form. These logical conditions are listed in Appendix \ref{sec: input combinations}. Furthermore, Figure~\ref{fig:input-condition-trees} illustrates the decision trees of the five most complex phenotype conditions, using the \emph{BN Classifier} tool~\cite{benevs2024bnclassifier}. This analysis yields significant interpretable insights into the phenotype repertoire of the Naldi et al model, as we describe next. Results such as these can be used as part of the model validation process and as novel model predictions.

We find that each input combination allows the coexistence of multiple dynamical phenotypes. For example, the absence of external TGFB ensures the coexistence of at least one dynamical phenotype corresponding to Th1 cells with at least one dynamical phenotype corresponding to Th2 cells.
As shown in the top left panel of Figure~\ref{fig:phenotype-count-histograms}, there is a robust representation of the co-existence of 2 to 12 dynamical phenotypes, and a significant fraction of input combinations allows the coexistence 15 or more phenotypes. 

There is a significant range in the number of input combinations that allow the existence of each dynamical phenotype, as we illustrate on the top right panel of Figure~\ref{fig:phenotype-count-histograms}. At the low extreme, the dynamical phenotypes corresponding to active Treg and Treg-Th17 cells with oscillating TGFB expression are present for only 4 input combinations. At the high extreme are dynamical phenotypes that represent resting Th1-Th17 hybrid cells, anergic Th1-Th17 cells, and resting Th2-Th17 cells, each of which exists under 49.2$\%$ of input combinations. Nine dynamical phenotypes are present under more than 25$\%$ of input combinations. 

We find that certain inputs have a canalizing effect on dynamical phenotypes. The most canalizing input is APC, which signifies the effect of antigen-presenting cells on activating the T cell receptor. Six dynamical phenotypes are possible for APC=0, all of which are resting cells. Twenty-two dynamical phenotypes are possible under APC=1; these describe active or anergic cells.

The complex, many-to-many mapping between external inputs and dynamical phenotypes supports experimental findings of a continuum of CD4+ T cell states \cite{eizenberg2017diverse}. The results of our dynamical phenotype analysis suggest that both hybrid cells and mixtures of multiple cell types that coexist under the same environment contribute to this continuum. 

\section{Identifying alternative phenotype-determining nodes from the canalization of the Boolean functions} 
\label{sec:LDOI_phenotypes}

 Selecting phenotype-determining nodes may not be straightforward in some systems, calling for methods that use information from the model to propose a candidate PDN set. Even when a biologically motivated PDN set exists, one must assess the degree to which it specifies the dynamics of the system and consider whether additional PDNs are needed for a more complete characterization. We find that in the T cell differentiation model~\cite{naldi2010diversity} each dynamical phenotype determined by the 9 PDNs includes a fixed state of 17 additional nodes (on average) out of 30 nodes that could be fixed in the trap space. For example, in the dynamical phenotype that corresponds to anergic Th2 cells, only 6 nodes other than the PDNs have the same value over all input combinations that allow this phenotype. This result implies that in this model, ``anergic Th2 cell'' is a constellation of many system states that are consistent with this classification, as also noted in the original article. This incomplete specification suggests the possibility that a different set of phenotype-determining nodes may be a better characterization of the state of the whole network.

We propose that a set of nodes with many logic sufficiency relationships forms an informative PDN set. Logic sufficiency means that the sustained state of a potentially distant regulator node yields the sustained state of a target node \cite{maheshwari2017framework}, and is captured by the concept of logical domain of influence (LDOI) introduced by Yang and collaborators \cite{yang2018target}. We characterize PDN sets by the average size of their contradiction-free LDOI (see Section~\ref{sec:LDOI} of the Methods for more details). We develop a greedy algorithm to incrementally grow the set of PDNs by adding a candidate PDN that is estimated to complement well the contradiction-free LDOI of the existing PDNs. 

We apply this algorithm to the T cell differentiation model of Naldi et al.~\cite{naldi2010diversity} and obtain a set of 5 PDNs, namely IL4R\_b1, NFAT, STAT1, TBET, and proliferation. Note that NFAT and TBET also belong to the 9 biologically-motivated PDNs. The node IL4R\_b1 denotes the basal level of the IL4 receptor. STAT1 is a transcription factor activated by cytokine signaling (via a JAK-STAT pathway). The abstract node ``proliferation'' functions as a switch in the Naldi et al. model: its activation indicates commitment to cell proliferation, which in the model is needed for the expression of cytokines.

Using this set of 5 PDNs, we obtain 32 dynamical phenotypes across 82 dynamical phenotype subspaces (see also breakdown in Table~\ref{tab:phenotype-entropy-ldoi}). We find that these are more specific than the 30 biologically-motivated dynamical phenotypes: The 5 LDOI PDNs fix a mean of 21 and a median of 22.5 nodes, compared to the mean and median of 17 nodes fixed by the 9 biologically-motivated PDNs. We interpret this as an indication that the LDOI-based PDN set is a better representation of the overall network state than the biologically-motivated PDN set. This is especially remarkable given the much smaller size of the LDOI-based PDN set. Furthermore, there are significantly fewer LDOI-based dynamical phenotypes for each input combination than biologically-motivated phenotypes. As shown in the bottom left panel of Figure~\ref{fig:phenotype-count-histograms}, the vast majority of input combinations correspond to the coexistence of 2-4 LDOI phenotypes, and the maximum number of LDOI phenotypes allowed for the same input combination is 6. This result aligns better with the biological expectation that each cell type requires a specific input combination (e.g., that differentiation of Th0 cells into Th1 cells requires a pro-Th1 microenvironment).

We visually compare the biological PDN set and the LDOI-based PDN set by indicating two subgraphs of the influence graph of the T cell differentiation model. The subgraph on panel A focuses on the 7 PDNs exclusive to the biological PDN set (marked with red border) and the one in panel B centers on the 3 exclusively LDOI-based PDNs (marked with blue border). The 7 biological PDNs are connected by canalizing paths to 10 other nodes, while the 3 LDOI-based PDNs are connected by canalizing paths to 22 other nodes, 16 of which are exclusive to them. This enrichment in canalizing regulation supports the appropriateness of the LDOI-based PDN selection and explains the higher specificity of the resulting dynamical phenotypes.

\begin{figure}
    \centering    
    \includegraphics[width=1.0\linewidth]{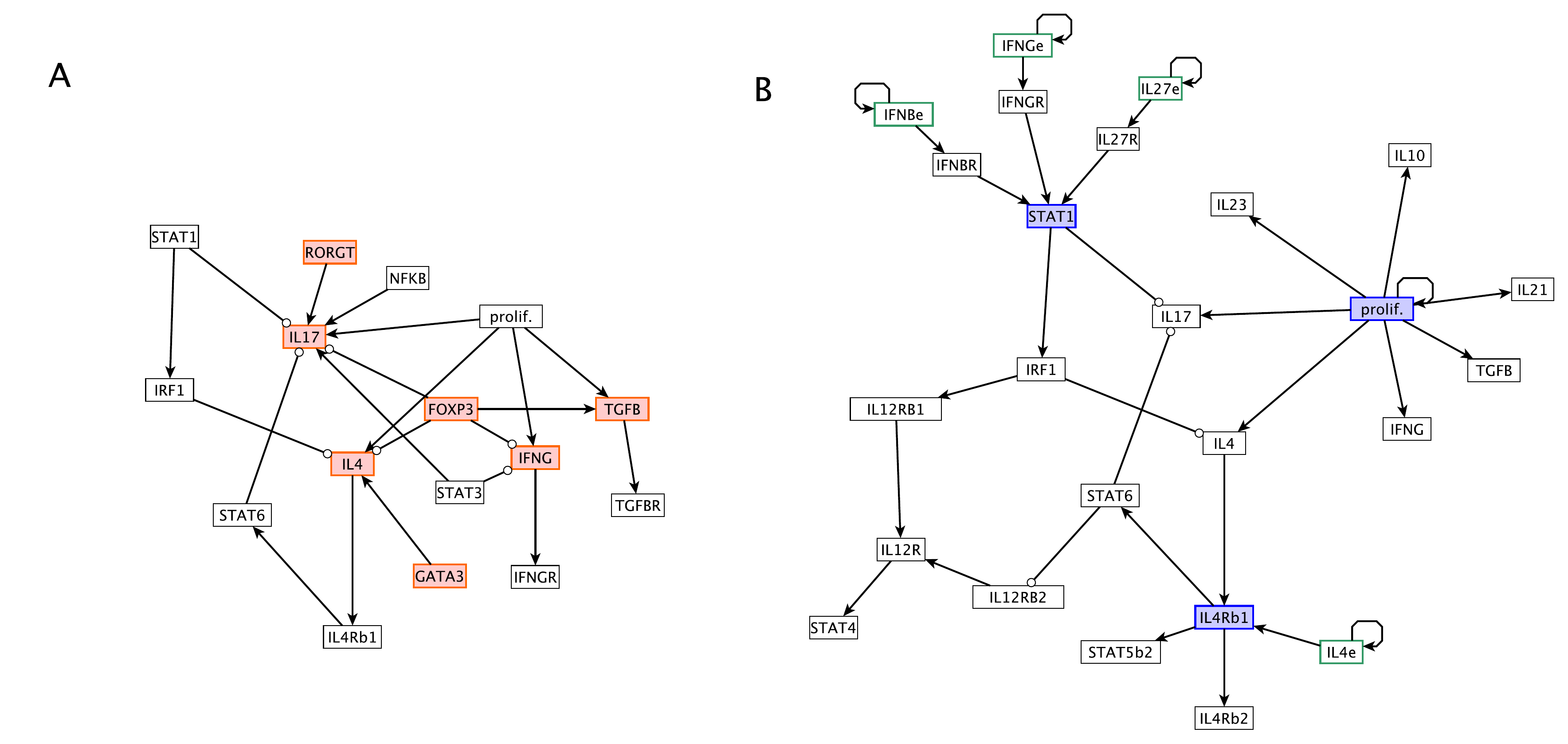}
    \caption{Two subgraphs of the influence graph of the T cell differentiation model, illustrating the canalizing regulation of PDNs and the canalizing influences incident on them.  Regular arrows represent positive influences (activation), and round arrows represent negative influences (inhibition). Panel A focuses on the 7 nodes that are exclusive to the biologically-motivated PDNs. The canalizing influences of these PDNs tend to remain within the set.  Panel B focuses on the 3 PDNs exclusive to the LDOI-based PDN set.  The canalizing influences of these PDNs extend to 14 nodes, including 4 of the 9 biologically-motivated PDNs. The LDOI-based PDNs are connected to 22 nodes by canalizing paths. The nodes with green outlines are external inputs.}
    \label{PDN_influences}
\end{figure}

Our analysis indicates that the LDOI-based dynamical phenotypes characterize the system state in a manner that is complementary to the information given by the biologically-motivated phenotypes. For example, the LDOI-based dynamical phenotypes reveal the strong influence of IL4R\_b1 and STAT1 on the cell state. The dynamical phenotypes with the same combination of IL4R\_b1 and STAT1 values have significant similarities. They appear under identical or near-identical input combinations and share the fixed state of 7 to 10 nodes (see Table \ref{LDOI_families}). Each combination of IL4R\_b1 and STAT1 values has a unique pattern of activation of four STAT nodes (STAT1, STAT4, STAT5 and STAT6). The article by Naldi et al. indicates that each of these STAT nodes is a compressed representation of a JAK-STAT pathway. Thus, the activity of a STAT node indicates biological information that can be extracted from the model and goes beyond the specific questions asked in the original article. Examples of such information include the activity of a JAK protein or of a target gene of a STAT transcription factor. 

Another aspect in which the LDOI-based dynamical phenotypes offer complementary information is specifying novel settings under which the cell-specific cytokines are inactive. Each combination of IL4R\_b1 and STAT1 values yields the inactivation of IFNG, IL4 and/or IL17. Another robust way of inactivating these cytokines is the OFF state of the LDOI PDN proliferation. Both of these LDOI PDN settings are complementary to the expected setting under which there is no cytokine production, namely NFAT=0.

\begin{figure}
    \centering    
    \includegraphics[width=1.0\linewidth, trim={3cm 1.8cm 3cm 1.8cm}, clip]{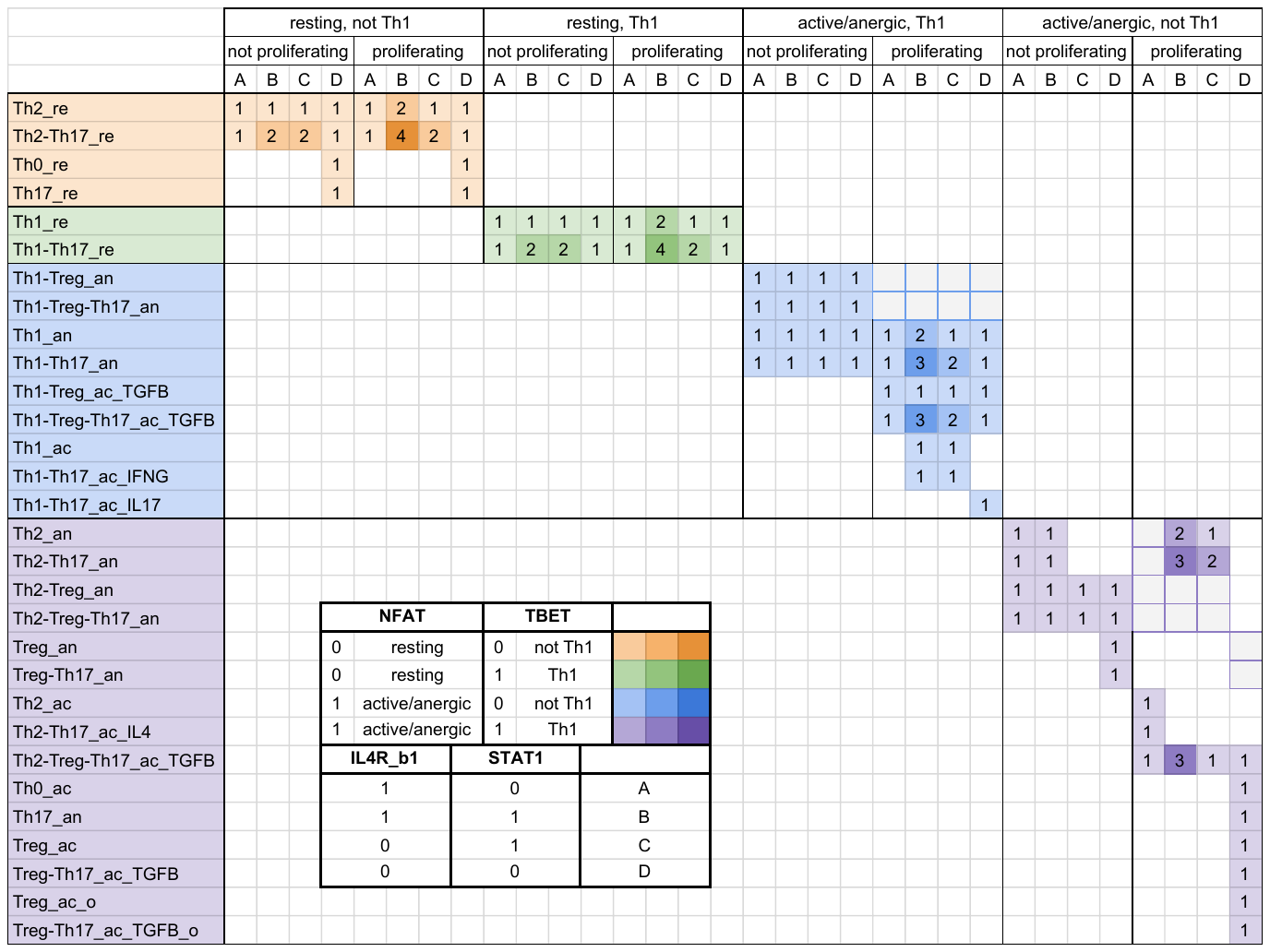}
    \caption{Mapping between the biologically-motivated dynamical phenotypes (rows, see Table~\ref{Bio_phenotypes} for the meaning of the notations) and the LDOI-based dynamical phenotypes (columns). The LDOI phenotypes are grouped according to the values of NFAT, TBET, proliferation, IL4R\_b1 and STAT1, as explained by the figure legend. The coloring is based on the values of NFAT and TBET (also see legend). Finally, the numbers and color intensity indicate the attractor count in thousands ($1 \equiv [1, 1000]$, $2 \equiv [1000, 1999]$, etc.). The matrix positions with a colored outline and grey background indicate intersections of biological and LDOI-based dynamical phenotypes that could but do not contain any attractors.}
    \label{fig:LDOIvsbiophenotype_matrix}
\end{figure}

To gain insight into the similarities and differences of the dynamical phenotypes corresponding to the two PDN sets, we partition the attractors into a matrix in which the biological phenotypes form the rows and the LDOI phenotypes form the columns. As shown in Figure~\ref{fig:LDOIvsbiophenotype_matrix}, we find significant regularities, which allow an almost block-diagonal organization of the non-zero entries in this matrix. A block-diagonal structure with small blocks would indicate a strong alignment of the dynamical phenotypes. In contrast, we find six blocks of significant size (with 16 or more nonzero entries), suggesting a weaker alignment. These blocks are determined by the combinations of the values of the shared PDNs NFAT and TBET, and of the LDOI-specific PDN proliferation. The value of NFAT distinguishes resting cells from active or anergic cells, the value of TBET separates Th1 cells and their hybrids from the rest of the cell types, and the value of proliferation distinguishes between active and anergic cells. The elongated shape of most blocks is a sign of differing resolutions between the biologically-motivated and LDOI-based dynamical phenotypes. The LDOI-based dynamical phenotypes have a higher resolution for resting cells, while the biologically-motivated phenotypes have a higher resolution for active/anergic cells with TBET=0. The most prominent divergence from block-diagonality is a proliferating phenotype with NFAT=1, TBET=IL4R\_b1=STAT1=0. This LDOI phenotype is consistent with 7 biological phenotypes that are either anergic, express TGFB, or have oscillating TGFB.

Around 17$\%$ of the intersecting dynamical phenotypes do not contain any attractors; these positions in the matrix are shown with colored outlines in Figure \ref{fig:LDOIvsbiophenotype_matrix}. The lack of representation is due to relationships among node states that are unspecified in a dynamical phenotype. The most frequent example is the relationship between the value of the transcription factor FOXP3 and the cytokine TGFB, both of which mark Treg cells. Both of these nodes have unspecified values in the proliferating LDOI phenotypes with NFAT=1. Yet, in these dynamical phenotypes the value of TGFB depends solely on FOXP3; this dependence selects the active Treg cell state and prevents the anergic cell state. If we were to run our LDOI algorithm further to reach 6 LDOI-based PDNs, the next addition would be FOXP3. The 6-PDN set would yield 48 dynamical phenotypes, doubling the number of LDOI-based dynamic phenotypes with NFAT=1. A benefit of this increased resolution is that it would reveal the lack of LDOI-based dynamic phenotypes in which FOXP3=proliferation=1 and TGFB=0.

We find that the number of attractors in the intersection of the two kinds of dynamic phenotypes mainly depends on the number of input combinations that allow both subspaces. For example, the intersections of the proliferative LDOI phenotype family with IL4R\_b1=STAT1=1 (which exists under 43.75\%  of input combinations) with resting Th1-Th17 cells or resting Th2-Th17 cells (both of which exist under 49.2\% of input combinations) have more than 3000 attractors.

\section{Validation of alternative PDN sets based on the attractor repertoire}
\label{sec:validation-of-pdns}

Our complete trap space-based definition of dynamical phenotypes is especially suited for systems in which it is not feasible or practical to determine the full attractor repertoire. The T cell differentiation model of Naldi et al.~\cite{naldi2010diversity} is nearing the limit of this possibility, having $2^{13}$ input combinations and 68,100 attractors. Since we were able to identify all the attractors, we can consider alternative definitions of phenotypes as attractor groups. We pursued two complementary attractor-based methods to identify PDNs and phenotypes. Comparing these phenotypes to the LDOI-based dynamical phenotypes serves as a test of the LDOI-based PDN identification method.

\subsection{Identifying phenotypes from grouping of attractors}
\label{sec:clustering-results}

We adapt a method often used to cluster single-cell RNA data~\cite{seth2022dimensionality, levine2015data} and apply it to the attractors of the T cell differentiation model. The methodological details are described in Section~\ref{sec:clustering}. In brief, we perform dimensionality reduction from the 58 Boolean variables using multiple correspondence analysis. We obtain a smaller set of 20 continuous variables that preserve 95\% of the attractor data variability. We then construct a weighted similarity graph from these continuous variables and identify the clusters of this graph using the Leiden bottom-up clustering algorithm. We obtain 31 attractor clusters. Finally, to identify the most effective PDN candidates, we rank model nodes based on how well they explain the classification of attractors into the identified clusters. To verify the robustness of this clustering approach, we validate the results across a range of parameters (further details are given in Section~\ref{sec:clustering}).

\begin{figure}
    \centering    
    \includegraphics[width=1.0\linewidth, trim={2.5cm 1.8cm 2.5cm 1.8cm}, clip]{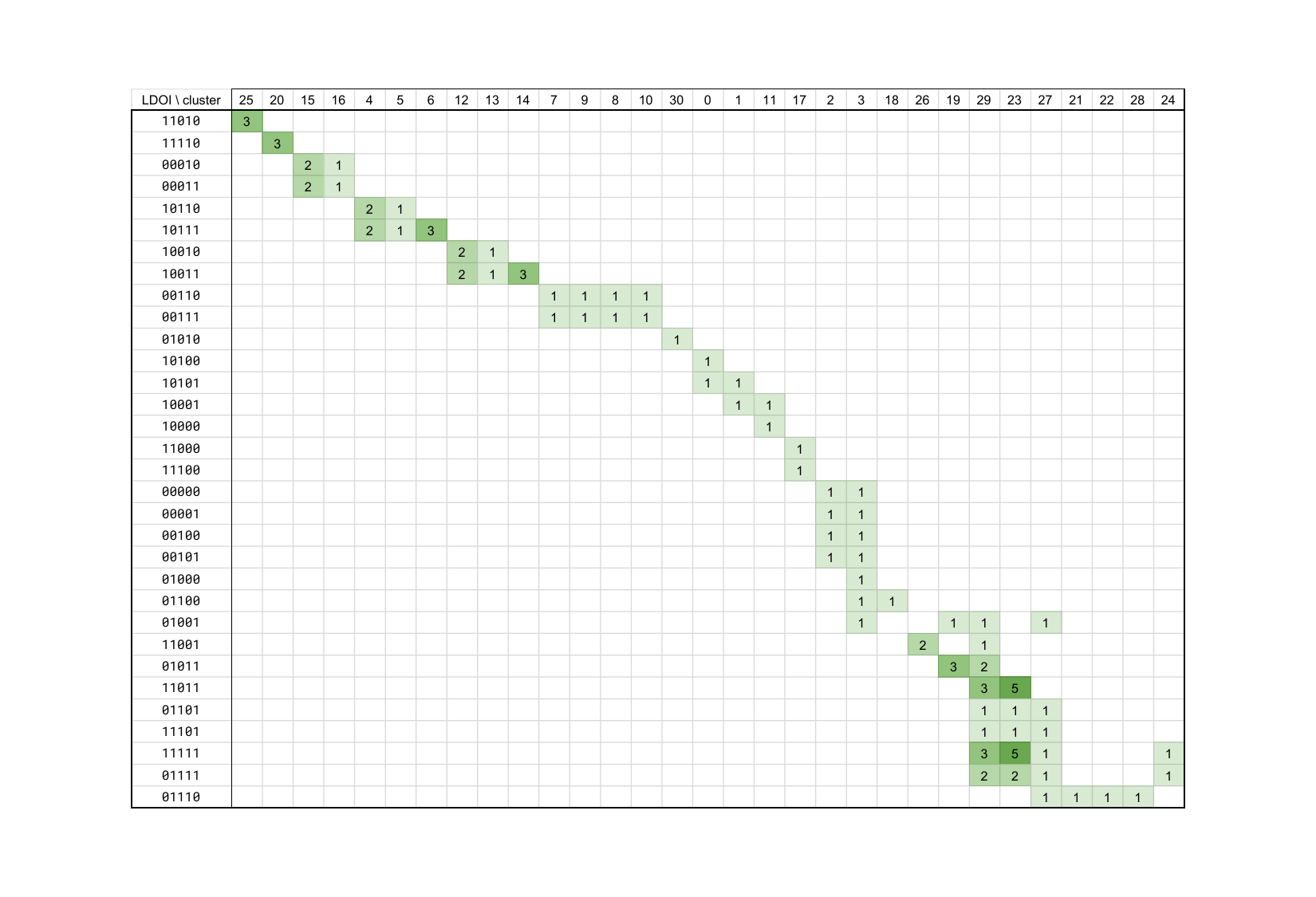}
    \caption{Mapping between the 32 LDOI-based phenotypes (rows) and the 31 attractor clusters (columns). The LDOI phenotypes are denoted by a binary string indicating the state of IL4R\_b1, NFAT, STAT1, TBET, and proliferation, respectively. The numbers and color intensity indicate the attractor count in thousands ($1 \equiv [1, 1000]$, $2 \equiv [1000, 1999]$, etc.).}
    \label{fig:clustervsLDOIphenotype_matrix}
\end{figure}

As illustrated in the mapping shown in Figure~\ref{fig:clustervsLDOIphenotype_matrix}, we find a significant agreement between the attractor clusters and the LDOI-based dynamical phenotypes. The eight rows with a single entry indicate that each of these eight LDOI phenotypes is enclosed within a single attractor cluster. The ten columns with a single entry indicate that each of these ten attractor clusters is enclosed in a single LDOI-based dynamical phenotype. There are 72 nonzero entries in this $32 \times 31$ matrix. Correspondingly, the average number of clusters overlapping an LDOI phenotype is 2.25, and the average number of LDOI phenotypes overlapping an attractor cluster is 2.32.  The strongest case of degeneracy is cluster 29, which overlaps 8 LDOI phenotypes.

To put these results into perspective, we determine the analogous mapping between the attractor clusters and the biologically-motivated dynamical phenotypes. As indicated in Figure~\ref{fig:clustervsbiophenotype_matrix}, their correspondence is weaker.
There are 100 entries in this $30 \times 31$ matrix, which implies that the average number of clusters overlapping a biologically-motivated phenotype is 3.33, and the average number of biologically-motivated phenotypes overlapping an attractor cluster is 3.22. Cluster 3 overlaps 14 biological phenotypes, and the dynamical phenotypes corresponding to Th1 cells and Th1-Th17 hybrids overlap 11 clusters each. 

Overall, the more focused correspondence of the attractor clusters with the LDOI-based dynamical phenotypes compared to the biologically-motivated dynamical phenotypes supports the appropriateness of the LDOI-based PDN identification.

\subsection{Identifying phenotype-determining nodes using mutual information}

To further evaluate the proposed LDOI method for identifying phenotype-determining nodes (PDNs), we compare it with an alternative approach that leverages information-theoretic measures applied to the model's attractors. It identifies PDNs by maximizing the mutual information between a candidate set of nodes and the remainder of the network. Mutual information quantifies the amount of information shared between two sets of nodes; intuitively, it indicates how well the states of one set can predict the states of the other. We employ a greedy algorithm that iteratively constructs the PDN set: it first selects the node with the highest mutual information relative to the rest of the network, then sequentially adds nodes that maximize the incremental mutual information with the remaining nodes, until the desired set size is reached. We find that the PDNs identified by this mutual information-based method exhibit substantial overlap with those obtained via the LDOI approach.

Table \ref{PDNcomparison} compares the PDN set identified by the LDOI method with the 8 top-ranked PDNs obtained by attractor clustering (Section~\ref{sec:clustering-results}) and mutual information. Both attractor-based methods  may identify candidate PDNs with the same score. In general, the choice of PDN at a given iteration affects which candidate PDNs are highly ranked in the next iteration. In the example of the T cell differentiation model of~\cite{naldi2010diversity}, however, for the first 8 iterations, the highest-scoring PDNs in each iteration are equivalent and do not affect which nodes are highly ranked in the next iteration. The candidate PDNs are equivalent choices because they form linear pathways (such as STAT1$\rightarrow$IRF1$\rightarrow$IL12RB1) or are co-regulated (e.g., NFAT and IL2RA are completely determined by APC). We note that although the table shows nodes at various positions, the interpretation should not compare the ranking of individual nodes, but rather the composition of the PDN sets that also include all nodes identified in previous steps. The reason for this interpretation is that the choice of a node as a PDN affects the appropriateness of the remaining nodes as candidate PDNs for each of the three methods.

First, we observe that the PDN sets of the two attractor-based methods align very well: the respective top-8-PDN sets share 7 nodes. Second, we find that the results of the attractor-based methods support the LDOI-based PDNs set. Four of the five LDOI PDNs are part of the top-4-PDN set according to attractor clustering and are part of the top-6-PDN set according to mutual information. 
The fifth LDOI PDN, proliferation, is selected later, in the 9th iteration for attractor clustering and in the 10th iteration in the mutual information-based method. This is likely because the setting under which the value of proliferation has a canalizing role (namely, NFAT=1, proliferation=0) appears in relatively few attractors (compare the right and left halves of the blue and purple blocks in Figure~\ref{fig:LDOIvsbiophenotype_matrix}).

Overall, the agreement between the PDN sets of the three methods and the alignment of the LDOI-based dynamical phenotypes and the attractor clusters provide validation that the LDOI method identifies nodes that reliably predict the attractors.

\begin{table}[]    
    \setlength{\tabcolsep}{10pt} 
    \renewcommand{\arraystretch}{1.5} 
    \normalsize
    
    \begin{tabular}{C{1cm}L{1.5cm}L{2.95cm}L{2.95cm}}
     Rank & LDOI & clustering-based & mutual information\\\hline\hline
     
     1 & \textcolor{blue}{IL4R\_b1} & \textcolor{blue}{IL4R\_b1}, STAT6, IL12RB2 & \textcolor{purple}{TBET}, RUNX\\\hline

     2 & \textcolor{red}{NFAT} & \textcolor{teal}{STAT1}, IRF1, IL12RB1 &  \textcolor{red}{NFAT}, IL2RA, IKB\\\hline

     3 & \textcolor{purple}{TBET} & \textcolor{red}{NFAT}, IL2RA, IKB & TGFBR, SMAD3\\\hline

     4 & \textcolor{teal}{STAT1} & \textcolor{purple}{TBET}, RUNX & \textcolor{blue}{IL4R\_b1}, STAT6, IL12RB2\\\hline

     5 &  proliferation &  SMAD3, TGFBR & IL4RA, STAT5\_b2\\\hline

     6 & & IL4RA, STAT5\_b2 & \textcolor{teal}{STAT1}, IRF1, IL12RB1\\\hline

     7 & & IL12R, STAT4 & IL23R\\\hline

      8 & & FOXP3, NFKB & FOXP3, NFKB\\\hline
     
    \end{tabular}
    \vspace{8pt}
    
    \caption{Top-ranked PDNs according to the LDOI method and two attractor-based methods applied to the T cell differentiation model. Multiple node names in the same position for the clustering-based and mutual information-based analysis of the attractor repertoire indicate that those nodes are equivalent PDNs. We mark the 4 PDNs identified by all three analysis methods with separate colors and highlight their occurrence in the top-ranked PDN sets.}
    \label{PDNcomparison}
\end{table}


\section{Discussion}
\label{sec:discussion}

Boolean models of biological systems often characterize the model outcomes by the states of a small set of output nodes (biomarkers). However, no theoretical framework has previously addressed how the choice of biomarkers impacts the mapping between model inputs and phenotypic outcomes.
Here we address this gap by defining dynamical phenotypes as complete trap spaces in which a set of designated biomarkers (or PDNs) have the most specific state possible, while input nodes have the least specific state possible. This definition allows the precise identification of the external input combinations that make each biomarker configuration dynamically achievable. 
Rather than viewing phenotypes as fully specified system states, dynamical phenotypes capture a commitment of the network dynamics to a constrained region of state space. This distinction between commitment and full specification naturally accounts for phenotypic stability in the presence of molecular noise. Complete trap spaces thus formalize the idea that biological phenotypes can be robust without being rigid, and stable without corresponding to a single attractor. As an added benefit, the complete trap spaces of a Boolean model are independent of the update scheme chosen (e.g., whether nodes are updated synchronously or asynchronously). Thus, our notion of dynamical phenotype is naturally applicable to models constructed for any update scheme.

Our binary decision diagram- based approach enables scalable identification of dynamical phenotypes without full attractor enumeration and yields interpretable input–phenotype mappings. Our case study demonstrates that this representation allows analyses that not only recapitulate known helper T cell phenotypes but also provide additional information with explanatory and predictive power. For example, we find that most external input combinations allow the coexistence of multiple CD4+ T cell types. Our input-to-phenotype mapping analysis allows the identification of environments that permit a specific mixture of cell types. The coexistence of multiple dynamical phenotypes under the same external conditions offers a natural dynamical explanation for experimentally observed population-level heterogeneity~\cite{eizenberg2017diverse}.

The ability to efficiently compute dynamical phenotypes for alternative PDN sets enables the evaluation of multiple biomarker choices and the selection of the most suitable PDN set. In this study, we evaluated a purely model-dependent method for PDN identification and found that the resulting dynamical phenotypes provide information complementary to the biologically-motivated phenotypes. This complementarity has practical implications: our analysis suggests that adding a biomarker of proliferation, STAT1, and/or IL4R could discriminate among resting T cell states that are not distinguishable by the 9 phenotype markers proposed in the original study. 

We further propose that identifying the number of attractors in the intersections of dynamical phenotypes corresponding to alternative PDN sets, as illustrated in Figure~\ref{fig:LDOIvsbiophenotype_matrix}, is an effective comparative analysis. Deviations of the resulting intersection matrix from a diagonal structure reveal the differences in the phenotypic resolution between PDN definitions. Larger blocks indicate weaker discriminatory power, while elongated blocks reflect asymmetries in resolution between the two types of dynamical phenotypes.  Attractor-free regions within a block indicate cell states possible under one dynamical phenotype definition but not under the other. Long tails that contain a considerable number of attractors indicate insufficient resolution and motivate the inclusion of additional PDN.

Historically, experimental biomarkers have been chosen based on what is readily observable. When many outputs are observable, clustering and enrichment have been used to identify phenotypes, but such purely data-driven methods miss the opportunity to leverage mechanistic understanding encoded in a model. At the other end of the spectrum, many modelers designate abstract phenotype nodes or manually specify phenotypes. Our results suggest that PDN selection should be treated as a decision that can be systematically explored, rather than fixed a priori. In practice, this decision will involve biological expertise together with optimization criteria. We recommend starting with biologically motivated PDNs. If the resulting phenotypes appear incomplete, one should supplement the biologically motivated PDNs with nodes that have a large canalizing influence on the network. We provide an LDOI-based method for identifying such candidate nodes. Biological relevance, experimental accessibility, and interpretability can be used to exclude nodes a priori. In our case study, we excluded external molecules that serve as inputs and receptors solely regulated by external molecules. Biological and practical considerations could also affect the metrics used for ranking candidate PDNs. For example, the size of the LDOI could be replaced by the size of the ``relevant LDOI'', which excludes certain nodes that aren't of interest.

The concept of dynamical phenotype also enables a reinterpretation of phenotype control as the external control of a minimal set of nodes that drives the system into the trap space that corresponds to the dynamical phenotype. Because such control does not require driving the system into a specific attractor within the trap space, it is less extensive and potentially easier to implement than attractor control \cite{rozum2022pystablemotifs}. The systematic evaluation of phenotype control interventions will be the topic of future work.

 The concept of dynamical phenotype may be extended to multi-level discrete models and continuous (ODE-based) models. The foundational concept of stable motif, which underlies trap spaces, has been extended to multi-level models~\cite{Gan2018general} and to ODE models~\cite{rozum_identifying_2018}, and methods to identify the corresponding trap spaces (positive invariant sets) exist~\cite{rozum_identifying_2018,Gan2018general}. The trap spaces of a Boolean model are also related to ODEs through the concept of multivalued refinements of a Boolean model~\cite{pauleve2020reconciling}. A multivalued refinement formalizes the notion of adding intermediate states between the binary on (maximum value) and off (minimum value) states. It has been shown~\cite{pauleve2020reconciling} that a Boolean update scheme involving intermediate ``increasing'' and ``decreasing'' states preserves transitions between maximal and minimal states for multivalued refinements of arbitrary resolution, including all ODEs that have derivative structures compatible with a given Boolean model. This Boolean update scheme also guarantees that the attractors and minimal trap spaces are the same. Therefore, there is a strong connection between the trap space structure of Boolean models and the possible quantitative refinements of the dynamics they describe. 
This connection presents an opportunity for improved efficiency when analyzing complex dynamical systems. The large state space of these models poses challenges to the identification of their attractor repertoire, yet the identification of the repertoire of dynamical phenotypes in a Boolean model is a more achievable goal. 

 Taken together, the phenotype-determining nodes and dynamical phenotypes identified by our methods yield testable predictions of informative biomarkers and phenotype-driving environments. These predictions enable principled prioritization of experiments. 
More broadly, this framework reframes phenotypes as commitments of network dynamics rather than fixed end states, providing a flexible and predictive lens for model-guided phenotyping and experimental design in complex biological systems.




\section{Methods}
\label{sec:methods}

In this section, we describe in detail our methods to identify dynamical phenotypes and phenotype-determining nodes.

\subsection{Symbolic identification of dynamical phenotypes}
\label{phenotype_subspace_determination}

Our definition of a \emph{dynamical phenotype} (Section~\ref{sec:dynamical-phenotype}) offers a naive way of detecting dynamical phenotype subspaces through enumeration: Assuming we can enumerate all complete trap spaces, we can also detect the ones wherein PDNs stay fixed across all smaller trap subspaces. Out of these, we can then identify those that are maximal in terms of non-PDN nodes. It is then trivial to merge these dynamical phenotype subspaces into the dynamical phenotypes. However, this naive method scales with the number of complete trap spaces, which is often exponential in the number of nodes $n$ ($3^n$ in the worst case).

Here, we propose a symbolic method based on \emph{binary decision diagrams} (BDDs)~\cite{Bryant86}, which does not scale with the absolute number of complete trap spaces, but rather with the number of unique combinations of PDNs ($3^{|\text{PDN}|}$ in the worst case; here, $|\text{PDN}|$ is the number of PDNs). The method still considers the search space of all complete trap spaces, but instead of enumeration, it represents the set symbolically using a BDD. This BDD can be computed in $\mathcal{O}(n)$ BDD operations, and each operation is in the worst case quadratic w.r.t. BDD size. Consequently, our method is a heuristic that significantly outperforms enumeration in situations where the investigated sets have a sufficiently compact BDD-based encoding (i.e., the BDD is significantly smaller than the direct representation of all its elements). Previous analyses of ensembles of Boolean models of biological systems~\cite{benevs2022aeon,trinh2025mapping} indicate that real-world Boolean models often have this property, and their features can be compactly represented using BDDs.

With this in mind, our method operates as follows:

\begin{enumerate}    
    \item We use the symbolic algorithms of~\cite{pastva2026} to compute the BDD-represented set of \emph{complete} trap spaces $C$ (using $\mathcal{O}(n)$ BDD operations) and \emph{minimal} trap spaces $M$ (using $\mathcal{O}(n^2)$ BDD operations).
    \item Using projection, we \emph{enumerate} the set $P$ of all unique PDN combinations that are present in $M$. Note that this set has at most $3^{|\text{PDN}|}$ elements.
    \item Then, for each $p \in P$, we compute its dynamical phenotype subspaces as follows:
    \begin{enumerate}
        \item Compute the BDD-represented sets $C_p = \{ s \in M \mid \forall x \in \text{PDN}.~s(x) = p(x) \}$ and $N_p = C \setminus C_p$. Intuitively, these are the subsets of $C$ that either match the phenotype description $p$ or not. For a given $p$, this restriction can be performed as a single BDD operation.
        \item The dynamical phenotype subspaces of $p$ are then computed as $D_p = \mathit{maximize}(C_p \setminus \mathit{superspaces}(N_p))$, i.e., the inclusion maximal complete trap spaces such that they do not contain a complete trap space with different PDN values. Here, $\mathit{maximize}$ (resp. $\mathit{superspaces}$) can be computed using $\mathcal{O}(n^2)$ (resp. $\mathcal{O}(n)$) BDD operations, as described in~\cite{pastva2026}. Note that the reference describes an analogous $\mathit{minimize}$ process, but the computation of $\mathit{maximize}$ is symmetric.
    \end{enumerate}
\end{enumerate}

Note that the resulting sets $D_p$ are still represented symbolically using BDDs. At this point, we can use $D_p$ to enumerate all dynamical phenotype subspaces matching $p$, whose count is typically much smaller than that of all complete or minimal trap spaces. In our T cell case study, the model has $\sim$68,100 minimal trap spaces, but only 145 dynamical phenotype subspaces when using the biologically-motivated PDNs, and 82 dynamical phenotype subspaces when using LDOI-based PDNs. Alternatively, we can query $D_p$ directly using symbolic set operations to explore its structure (if the number of dynamical phenotype subspaces is still very large). Also note that if we are only interested in a single fixed combination of PDN values $p$, we may benefit from computing $C_p$ and $N_p$ directly, instead of requiring the knowledge of the full set $C$.

Our method is implemented in Python, using the symbolic trap space representations developed within the tool \texttt{aeon}~\cite{benevs2022aeon}. Its results can then be integrated with the tool \texttt{biobalm}~\cite{trinh2025mapping}, where it can be used to visualize the dynamical phenotypes within the succession diagram of the Boolean model.

\paragraph{Input-phenotype map}

To analyze dynamical phenotypes, we are often interested in their dependence on model inputs, i.e., in the minimal combinations of model inputs that need to be fixed to allow a specific combination of PDN values $p$. This information can be easily extracted from the BDD representation of the $D_p$ sets by projecting them to the model input nodes. The projected BDD can then be translated into logical conditions on the input nodes that guarantee the presence of a dynamical phenotype. We express these conditions in disjunctive normal form; they can even be made minimal by extracting their \emph{prime implicants}~\cite{rozum2022pystablemotifs}. For complex conditions or cases where we aim to compare the activity of multiple phenotypes, we can also consider decision trees and other forms of discrete bifurcation diagrams~\cite{benevs2024bnclassifier}.

We describe the results of our analysis of the input combinations corresponding to each dynamical phenotype of the T cell differentiation model in Section~\ref{sec:input-phenotype results}. A list of these input combinations is given in Appendix~\ref{sec: input combinations}. As an additional illustration, we take the five most complex phenotype conditions and visualize them as decision trees using the \emph{BN Classifier} tool~\cite{benevs2024bnclassifier}; the results are shown in Figure~\ref{fig:input-condition-trees}. 

We note that in the presence of multi-stability, fixing model inputs may not be sufficient to \emph{guarantee} that a specific PDN combination is eventually reached, as reachability is subject to the initial conditions and update scheme. In such cases, the set $D_p$ can also be used to derive such initial conditions for further analysis. Enumerating individual dynamical phenotype subspaces stored in $D_p$ yields complete trap spaces that, if reached, guarantee that the corresponding PDN values are sustained indefinitely. Each such maximal trap space therefore naturally identifies a minimal set of initial conditions for the model nodes that guarantee expression of a specific phenotype.

\subsection{LDOI-based identification of PDNs}
\label{sec:LDOI}

The \emph{logical domain of influence} (\ldoi) of a subspace $S$, denoted $\ldoi(S)$, is a subspace obtained by iteratively percolating $S$ through the model's update functions~\cite{yang2018target, rozum2021parity}. 
In particular, we iteratively construct sets $\ell_j$ beginning with $\ell_0=\emptyset$ and forming $\ell_{j+1}$ as the subspace given by $i\mapsto x_i$ for all $i$ for which 1) $f_i$ evaluates to $x_i$ everywhere on $\ell_j\cup S$ and 2) $i\mapsto \neg x_i$ is not fixed by $S$. For sufficiently large $j$, $\ell_j=\ell_{j+1}$ becomes constant; $\ldoi(S)$ is defined to be this limiting set.
Note that $S \subseteq \ldoi(S)$ is not guaranteed---for $(i \mapsto x_i) \in S$, we have $(i \mapsto x_i) \in \ldoi(S)$ only if the update function of $i$ evaluates to $x_i$ everywhere on the subspace $\ldoi(S) \cup S$. 
It is possible that $(i \mapsto x_i) \in S$ and $f_i$ evaluates to $\neg x_i$ everywhere on the subspace formed by $\ldoi(S)\cup S$. 
In such cases, $i \mapsto \neg x_i$ does not belong in $\ldoi(S)$, and instead it belongs in the \emph{contradiction boundary} of $S$~\cite{rozum2021parity}. We refer to subspaces with empty contradiction boundary as contradiction-free. If $S$ has a non-empty contradiction boundary, then it does not contain any trap space.

To identify potentially informative PDN sets, we introduce the term \emph{coverage}. Let $N$ be a subset of model nodes. Then the \emph{coverage} of $N$ is the average LDOI size across all contradiction-free subspaces $S$ that are created by fixing the values of all model nodes in $N$. Here, we exclude subspaces with non-empty contradiction boundaries, because they can never correspond to a trap space, and by extension can never yield a dynamical phenotype. We propose the heuristic that the greater the coverage of $N$, the more effective $N$ is in driving the rest of the nodes towards fixed values, and such $N$ is then also more informative as a PDN set.

To identify a set of model nodes with high coverage (which, by our heuristic, is a suitable PDN set candidate), we follow a greedy process that gradually increases the size of the PDN set by adding eligible nodes from a predetermined pool. In this work, we take as eligible every node that is not an input node, or is not regulated directly and exclusively by an input node. However, other choices of eligible node sets are possible depending on the biological assumptions of the model. The addition of new PDNs continues until a stopping criterion is reached. One possible stopping criterion, which we used here, is that the increase in coverage no longer surpasses a given threshold.

The first PDN is the eligible node with the largest LDOI. We then follow a two-step process to greedily identify the next node that is added to the PDN set. The first step aims to reduce the combinatorial complexity of the problem by computing a complementarity score that estimates the potential contribution of each candidate $i$ to the coverage of the current PDN set. In the second step, we then exactly compare the fitness of the best-performing candidates.

\begin{enumerate}
    \item We compute the complementarity score for each node $i$ with respect to the current candidate PDN set $N$. 
    \begin{enumerate}
        \item We identify partially consistent node assignments $i \mapsto x_i$ for each contradiction-free subspace $S$ corresponding to $N$ as those satisfying $\ldoi(\{ i \mapsto x_i \})\cap S\not=\emptyset$ and $\ldoi(S)\cap\{ i \mapsto x_i \}\not=\emptyset$. Note that two subspaces have an empty intersection only if there exists a node that is fixed in two opposite states in the two subspaces. 
        \item For each $i$ with at least one partially consistent assignment, we compute a complementarity score as the average of $|\ldoi(\{i \mapsto x_i\}) \cup \ldoi(S)|-|\ldoi(S)|$ across the partially consistent choices of $i \mapsto x_i$ and $S$ for the current $N$.
    \end{enumerate}
    \item The nodes with the top $k$ complementarity scores become finalist PDNs for the second step. Here, we exactly calculate the coverage of the union of the current PDN set with each of these $k$ finalist PDNs. The finalist PDN that yields the set with the largest coverage is adopted as the new PDN. 
\end{enumerate}

We note that computing the complementarity score for each PDN candidate $i$ requires only the LDOIs of its two states (which can be precomputed). The $2^{|N|}$ LDOIs defined by the current $N$ are also required, but they are carried over from the second step of the previous iteration. The $2^{|N|+1}$ LDOI calculations in the second step are done only for the $k$ finalist PDNs. The value of $k$ can be adapted to fit the specifics of each model. For example, in our analysis of the Naldi et al. model we set $k=5$ but using $k=3$ would have yielded identical PDNs (see below).

In our analysis of the Naldi et al. model, we considered $k=5$ finalist PDNs in the second step and used a coverage increase threshold of 2 for the stopping criterion. This means the process stops when adding the next best PDN only determines the state of the node itself and (on average) less than one additional model node. We found that in our case study, the greedy estimate in the first step was very good in finding the PDN finalists whose addition led to the maximal increase in PDN set coverage. In particular, when applying the algorithm to identify a 10-node PDN set, the maximal coverage increase corresponded to the PDN candidate with the highest or second-highest complementarity score for 9 of the 10 PDNs, and it corresponded to the PDN candidate with the third-highest complementarity score in the remaining case. The complementarity score slightly underestimated the coverage increase, but stayed within a difference of 0.75 nodes. For the Naldi et al. model, the first PDN has a coverage of 5, the increase in coverage is 5.5 when adding the second PDN, 2.3 when adding the 5th PDN, and drops to 1.5 when considering the 6th PDN. We thus obtained a PDN set of size 5. We then determined the dynamical phenotypes corresponding to this PDN set as described in Section~\ref{phenotype_subspace_determination}.

\subsection{Identification of phenotypes by attractor clustering}
\label{sec:clustering}

This method proposes to determine PDNs analogously to how biomarkers are commonly identified when studying single-cell genomic observations~\cite{seth2022dimensionality,levine2015data}. Here, we assume that individual states of our Boolean model correspond (on some qualitative level) to the possible cellular states of the biological system. As such, we treat each attractor as a cell observation taken in one of the biological phenotypes, as often done in the literature~\cite{wooten2019systems, cohen2015mathematical}.

The actual method then has to adapt to the underlying discrete data domain, but otherwise follows established best practices in dimensionality reduction, unsupervised clustering, and feature extraction for single-cell data:

\begin{enumerate}
    \item As the input for our method, we require the full set of model attractors. For complex attractors, we operate on the smallest enclosing subspace of the attractor, meaning the value of node $x_i$ in attractor $A$ can be $\{0, 1, \any\}$.
    \item As the first step, we take a subset of eligible nodes (same as in Section~\ref{sec:LDOI}; i.e., all nodes except for inputs and those regulated by a single input) and project the attractor set only to these eligible nodes.
    \item We perform dimensionality reduction of the discrete attractor observations onto a low-dimensional continuous space using \emph{multiple correspondence analysis} (MCA) from the \texttt{prince} Python package~\cite{Halford_Prince}. MCA is a discrete analog of \emph{principal component analysis} (PCA), the classical method for single-cell analysis. We only consider the first $n_{\text{MCA}}$ continuous dimensions, such that these dimensions cover at least $t_{\text{MCA}}\%$ of the variance of the original dataset. Here, $t_{\text{MCA}}\%$ is a method parameter.
    \item After MCA, we use unsupervised Louvain clustering~\cite{blondel2008fast} (as implemented by the Python package \texttt{scanpy}~\cite{wolf2018scanpy}) to identify clusters of closely related attractors. At this point, it is generally recommended to assess the stability of the clustering by testing different cluster resolutions or sub-samples of the original dataset~\cite{kamimoto2023dissecting}.
    \item We compute the Theil's U score (also known as uncertainty coefficient) between each model variable and the cluster labels using the \emph{mutual information score} (available in \texttt{scikit-learn} Python package~\cite{pedregosa2011scikit}). We select the best-scoring model variable as the first PDN. To select the subsequent PDNs, we follow the same approach, but instead of considering as baseline the entropy of the whole dataset, we consider a conditional entropy that accounts for the already chosen PDN node(s). That is, in the subsequent steps, we are searching for a model feature that best predicts the cluster labels, assuming the values of the already selected PDN nodes are known.
\end{enumerate}

This method yields clusters of closely related attractors based on the MCA dimensionality reduction, as well as the collection of candidate PDNs (sorted by importance) that best determine the partitioning of attractors into these clusters. However, it should be noted that the method is subject to several tunable parameters: in particular, the cutoff threshold for the MCA analysis and the selection of internal parameters of the Louvain clustering method. As such, it is generally recommended to evaluate the robustness of the result across a wider range of parameters. The method selects PDNs that are overall most likely to identify a specific cluster of attractor observations, but this may not cover all attractors across all clusters.

In our application to the attractors of the T cell differentiation model~\cite{naldi2010diversity}, the $\sim$68,100 attractors project to 3325 unique attractor observations when not considering inputs and nodes regulated by a single input. We set $t_{\text{MCA}}\%$ to $95\%$, resulting in $n_{\text{MCA}} = 20$. Subsequently, in Leiden clustering, the granularity of the final clusters is determined by the \emph{resolution} parameter and by an (optional) sub-sampling applied to the dataset. In our case, we selected a resolution of $0.4$, as it results in 31 attractor clusters, which closely matches the granularity of the phenotypes derived from the LDOI PDNs. To test the robustness of this choice, we compare the importance ranking of nodes using a wider range of resolutions. We observe that this ranking is largely stable, and in particular the set of top-five PDNs is the same for all resolutions (see Figure~\ref{fig:feature-importance-variations}; top). As an additional test, we simulate the impact of incomplete attractor knowledge on the node ranking by sampling subsets of the full attractor repertoire. Here, the impact is more pronounced, but for large sample sizes (e.g. $> 60\%$), the ranking is largely stable (see Figure~\ref{fig:feature-importance-variations}; bottom). These results demonstrate that Leiden clustering is able to capture the overall structure of discrete attractor data.

Finally, note that due to the nature of Leiden clustering, it is not guaranteed that the smallest subspaces enclosing each cluster are disjoint. This means that not every pair of clusters can be completely separated by fixing a subset of model variables. In our case, 7/465 ($\sim 1.5\%$) cluster pairs have intersecting subspaces; these intersections contain 3902 attractors ($\sim 5.7\%$ of all attractors). As such, while each attractor is assigned to exactly one cluster, it is not possible to find a PDN set that perfectly matches the results of the clustering. Still, we observe that for the majority of clusters (and attractors in them), the clustering is aligned with network subspaces.

\subsection{Identification of PDNs by mutual information}
This method is akin to a top-down analysis of the attractor repertoire. Here we assume that we have full knowledge of all the attractors. The selection of each PDN divides the attractors into separate clusters, where each cluster has either 0, 1, or oscillation for the value of the PDN. Our goal is to select PDNs such that each of their values best represents the attractors in each of the corresponding clusters.

One way to achieve this goal is through maximizing the mutual information between the PDNs and the rest of the nodes. The larger the mutual information, the better the PDNs are at predicting the values of the rest of the nodes, and therefore work better as biomarkers.

The information entropy indicates that the information gain from knowing the value of a PDN is determined by the possible values of that node across the attractors. The maximal information gain is when the node takes each of the 3 possible values evenly across all attractors, leading to $\log_2(3)$. The mutual information indicates the extent to which knowing the state of a set of PDNs yields information about the state of the rest of the nodes. The mutual information is limited by the information stored in either of the sets. Thus, the mutual information can increase steadily until roughly half of the nodes are selected as PDNs.

One can select any set of nodes as PDNs and calculate their mutual information with the rest of the nodes. However, as the number of possible sets grows intractably large as the size of the set grows, we use a greedy algorithm to select the PDN set. We start with the node with the highest mutual information relative to the rest of the network, then sequentially add nodes that maximize the gain in the set's mutual information with the remaining nodes, until the gain in mutual information decreases below a threshold. Considering every remaining node at each addition requires at most $N(N+1)/2$ computations of mutual information, making it economical. We exclude the source nodes and nodes directly and exclusively regulated by a source node from being a PDN and from the calculation of the mutual information.

In our analysis of the Naldi et al. model, we found that the mutual information increases monotonically as the PDN set grows. The gain in mutual information due to the first PDNs is close to 1, but the gain steadily decreases to less than 0.5 after 7 nodes and to less than 0.01 after 12 nodes.
As indicated in Table~\ref{PDNcomparison}, the PDNs identified by this method are in good agreement with those identified by attractor clustering.


\section*{Declarations}


\bmhead{Funding}
SP was supported by the MUNI/JS/1954/2025 project of Masaryk University.
KHP and RA were supported by the Pennsylvania State University.
Pacific Northwest National Laboratory is a multiprogram national laboratory operated by Battelle for the US Department of Energy under Contract No. DE-AC05-76RL01830.
VGT was supported by the Ho Chi Minh City University of Technology (HCMUT), VNU-HCM.

\bmhead{Competing interests} SP, KHP, JCR, VGT declare no financial or non-financial competing interests. RA serves as Associate Editor of this journal; she has no role in the peer-review of this manuscript or the decision to publish. RA declares no financial competing interests.

\bmhead{Ethics approval and consent to participate} Not applicable.

\bmhead{Consent for publication} Not applicable.

\bmhead{Data and code availability} All code and datasets necessary for reproducing the results of this study are available at \url{https://doi.org/10.5281/zenodo.18511140}.

\bmhead{Materials availability} Not applicable.

\bmhead{Author contributions}

RA and VGT conceived the project. SP and KHP developed the methodologies and corresponding software. All authors conducted the investigation and prepared the visualizations. SP, JCR, and RA wrote the original manuscript. All authors reviewed and edited the manuscript. 
All authors read and approved the final manuscript.








\bibliography{sn-bibliography}

\begin{appendices}

\section{Dynamical phenotypes agree with manual phenotyping}
\label{SI:three_examples}
Here we present examples of published models in which the attractors were reported for all input combinations. We demonstrate that a designation of phenotype-determining nodes exists that aligns groups of attractors, dynamical phenotypes as defined here, and biological phenotypes. 

1. The model by Cohen et al. \cite{cohen2015mathematical} describes the molecular pathways enabling cancer cell invasion and migration. The model has 32 nodes, of which DNAdamage and ECMicroenvironment are inputs. The phenotype marker/output nodes are Metastasis, Apoptosis, and CellCycleArrest, Migration, Invasion, and EMT. The original paper classifies attractors into four groups based on values of the output/phenotype nodes (homeostasis, Apop1 + Apop2 + Apop3 + Apop4, EMT1 + EMT2, M1 + M2). The four dynamical phenotypes are identical to these groups. The homeostasis phenotype is possible when both inputs are absent, the apoptosis phenotype is possible when DNAdamage=1, the EMT phenotype is possible when ECMicroenvironment=0, and the migration phenotype is possible when ECMicroenvironment=1.

2. The model by Wooten et al. \cite{wooten2021mathematical} describes the intracellular network that governs the transition of the opportunistic pathogen Candida albicans between two distinct phenotypes. The yeast form consists of single round cells, while the hyphal form consists of long, multicellular, branching tubular structures. C. albicans can also form an intermediate filamentous morphology called the pseudohyphal form. The model has 19 nodes, of which pH, Temperature, and Farnesol are input nodes. Three abstract nodes refer to processes:  hyphal\_initiation, hyphal\_maintenance, and HAG\_transcription. The original article reports 27 attractors, which are classified into four phenotypes: Yeast, Yeast-like (which corresponds to an initiated but then arrested transition), Hyphal-like (which is interpreted as a pseudohyphal form), and Hyphal. 

Designating the three abstract nodes as phenotype-determining nodes yields four dynamical phenotypes that are identical to the phenotypes reported in the paper. The Yeast phenotype is possible for 3 input combinations in which pH=0 and either Temperature =0 or Farnesol=1. The Yeast-like, Hyphal-like, and Hyphal phenotypes are possible for all input combinations.

3. The model by Dahlhaus et al. \cite{dahlhaus2016boolean} studies the role of the Aurora Kinase A in neuroblastoma by focusing on the mitosis part of the mammalian cell cycle. The network contains 23 nodes, including 4 inputs: AJUBA, GSK3B, MTCanAct, STMNCanAct. There is no sink node. Simulations with synchronous update yielded 50 attractors, which the authors classify into three cell behaviors: an oscillation corresponding to faithful mitosis, an oscillation that represents defective mitosis, and a fixed point that corresponds to the G0 checkpoint. The original article identifies the gene GWL/MASTL as an important determinant of the cell behavior: GWL/MASTL oscillates in the oscillatory attractors, GWL/MASTL=0 in the fixed point, and fixing GWL/MASL=1 increases the reachability of the normal mitosis attractors to the detriment of the other two types of attractors. Follow-up analysis of this model found that the oscillation that represents defective mitosis is an artifact of synchronous update, while the oscillation that represents faithful mitosis and the fixed point are preserved under asynchronous update~\cite{park2023models}.
 
 We found that designating GWL/MASTL as a PDN yields two dynamical phenotypes, one corresponding to faithful mitosis and the other to the G0 checkpoint identified in the original article. These dynamical phenotypes agree with the asynchronous attractors. Both dynamical phenotypes exist for any combination of the four inputs. 

\section{Input combinations supporting biologically-motivated dynamical phenotypes of the T cell differentiation model}
\label{sec: input combinations}

\begin{figure}
    \centering
    \begin{minipage}{0.3\linewidth}
        \centering
        \includegraphics[width=1.0\linewidth]{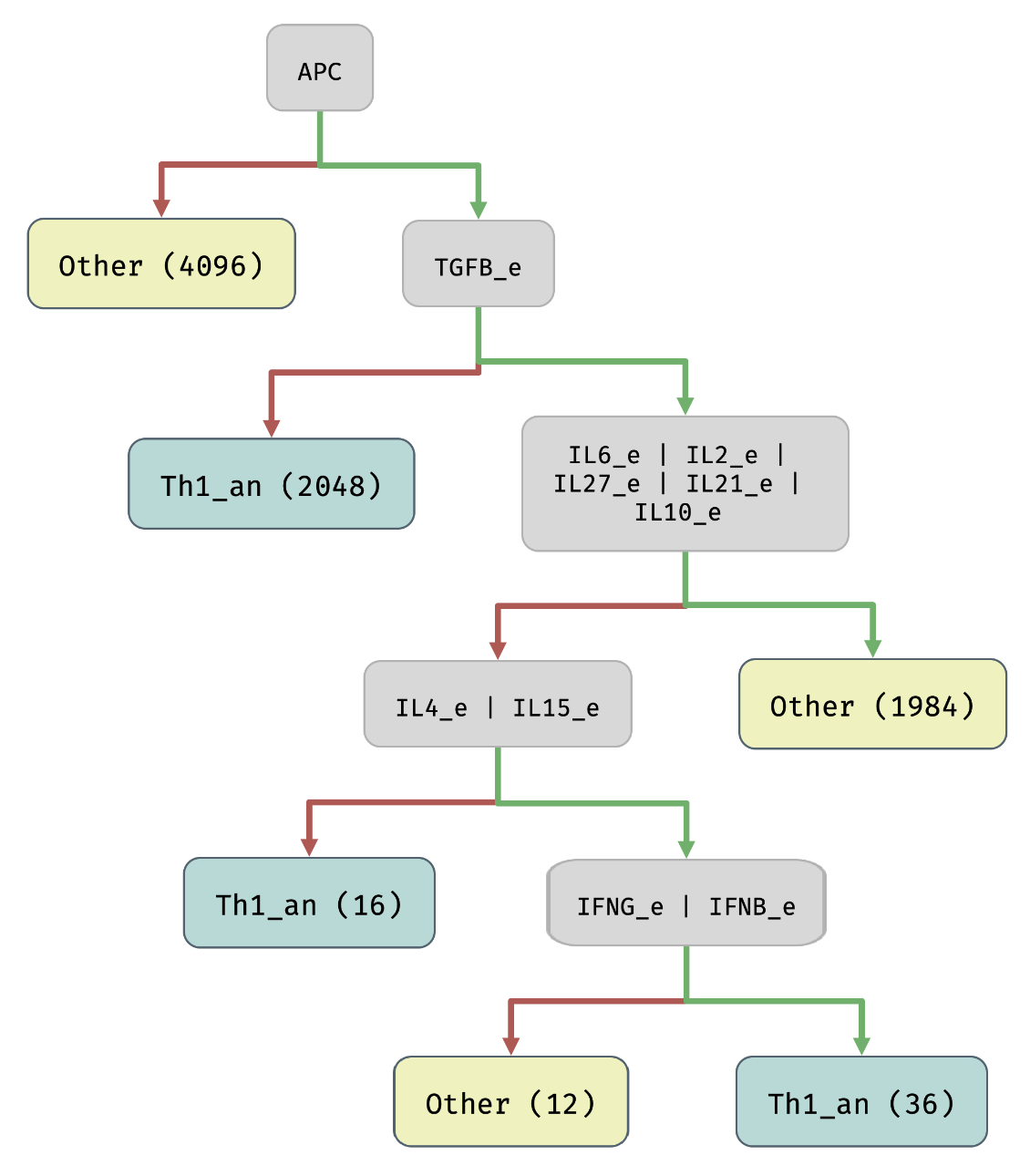}

        \vspace{3pt}
        (a) \texttt{Th1\_an}
    \end{minipage}
    \begin{minipage}{0.6\linewidth}
        \centering
        \includegraphics[width=0.9\linewidth]{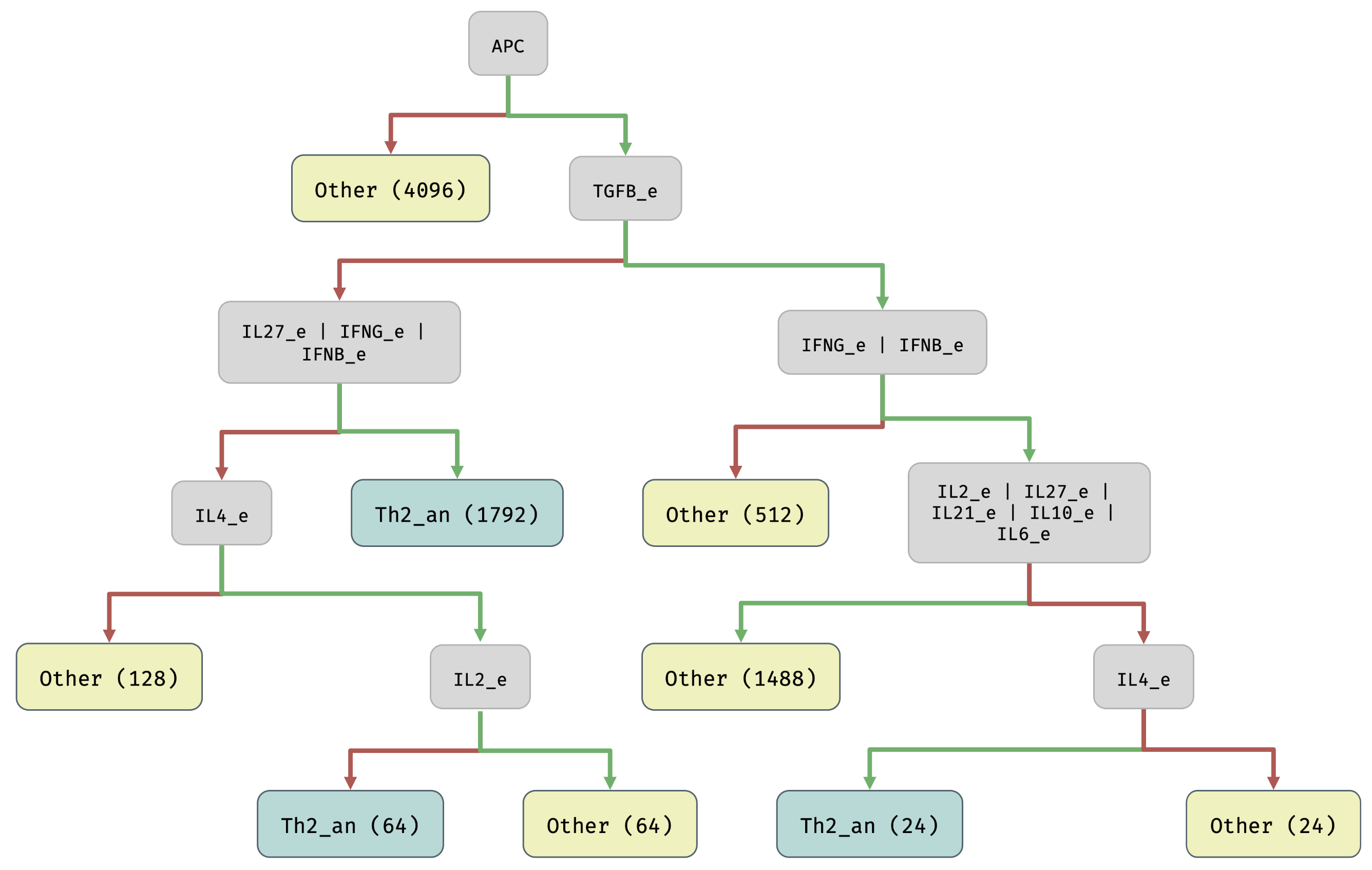}

        \vspace{3pt}
        (b) \texttt{Th2\_an}
    \end{minipage}

    \vspace{5pt}

    \begin{minipage}{0.48\linewidth}
        \centering
        \includegraphics[width=1.0\linewidth]{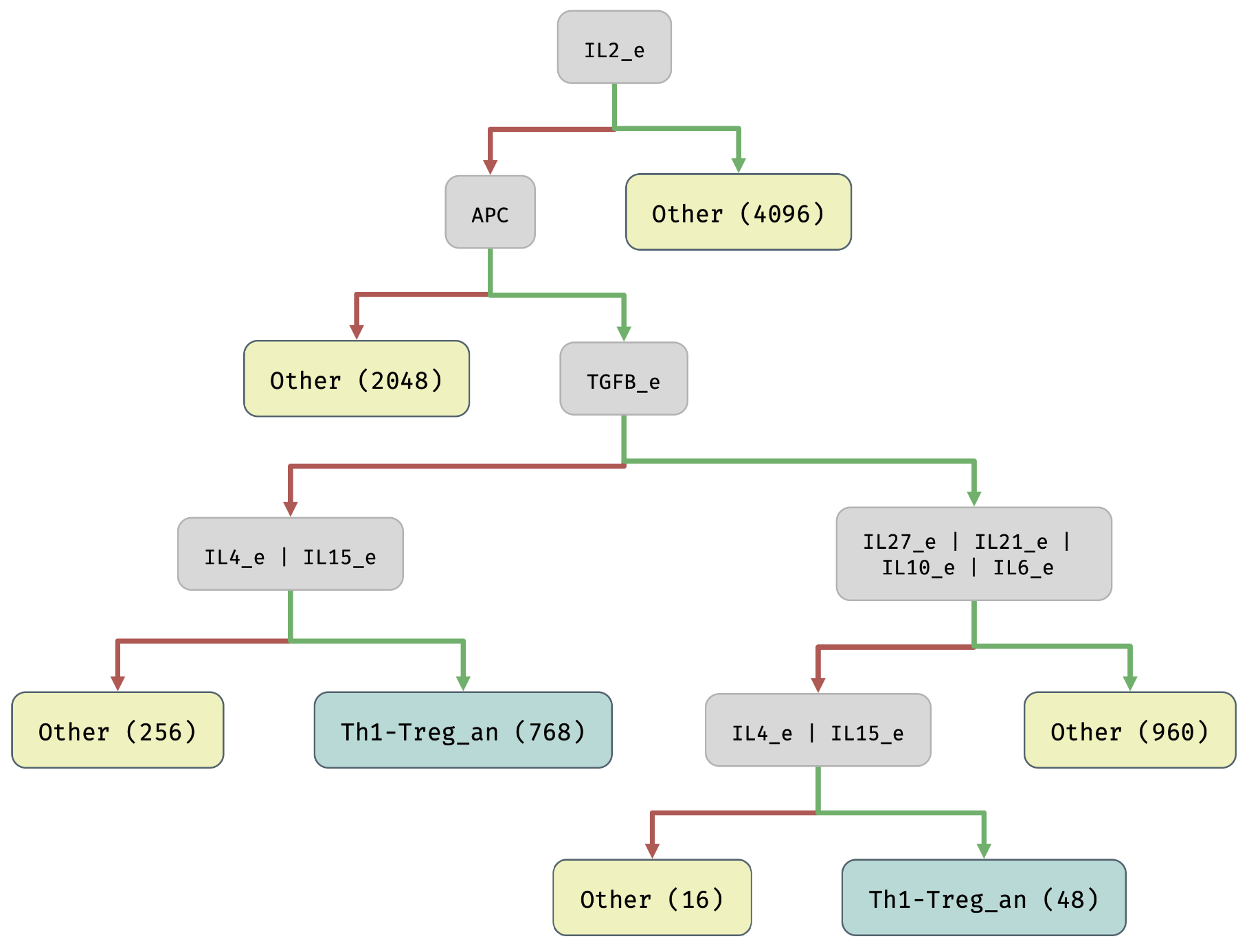}

        \vspace{3pt}
        (c) \texttt{Th1-Treg\_an}
    \end{minipage}
    \begin{minipage}{0.48\linewidth}
        \centering
        \includegraphics[width=0.5\linewidth]{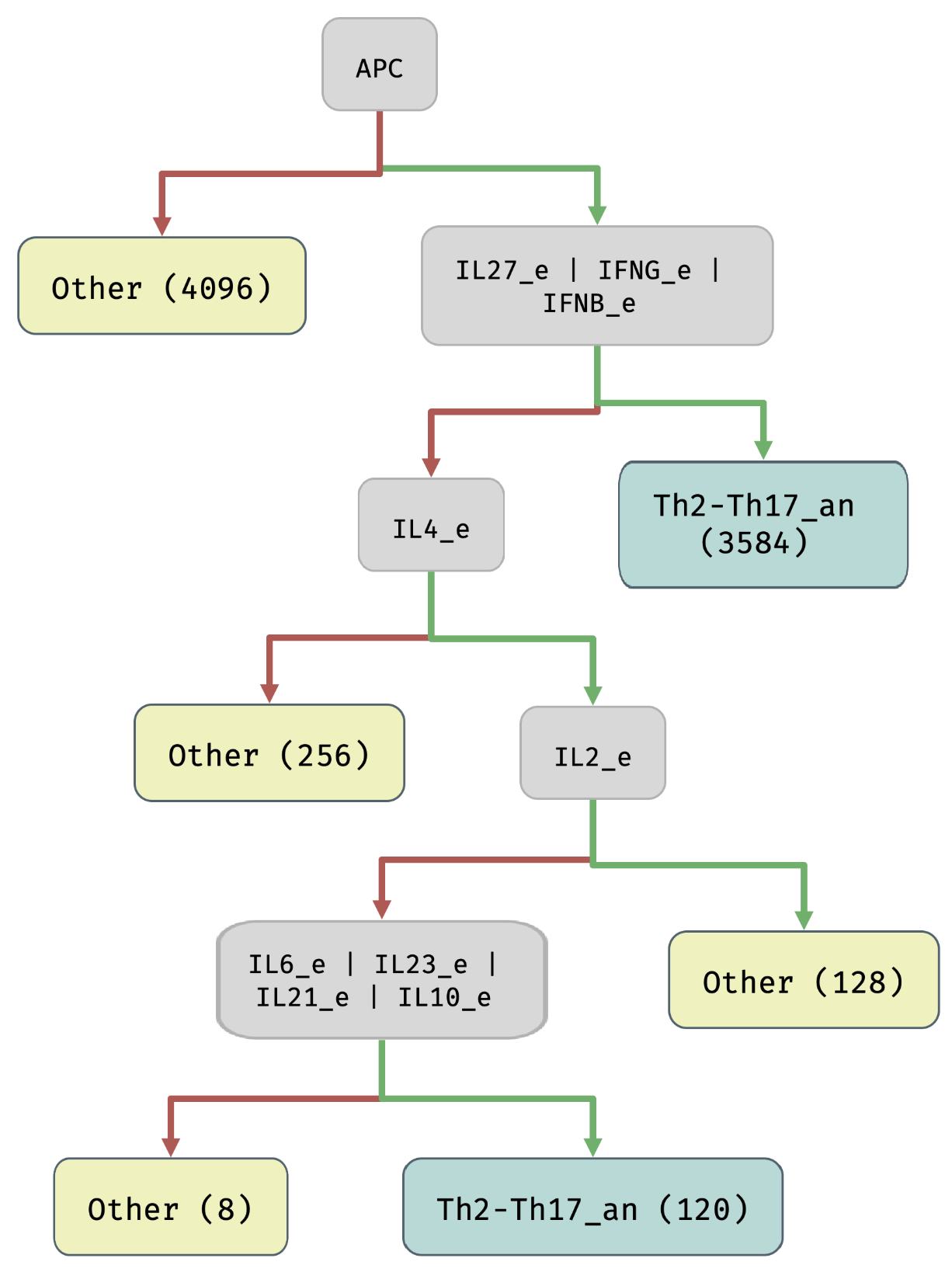}

        \vspace{3pt}
        (d) \texttt{Th1-Th17\_an}
    \end{minipage}

    \vspace{5pt}
    
    \begin{minipage}{0.9\linewidth}
        \centering
        \includegraphics[width=0.5\linewidth]{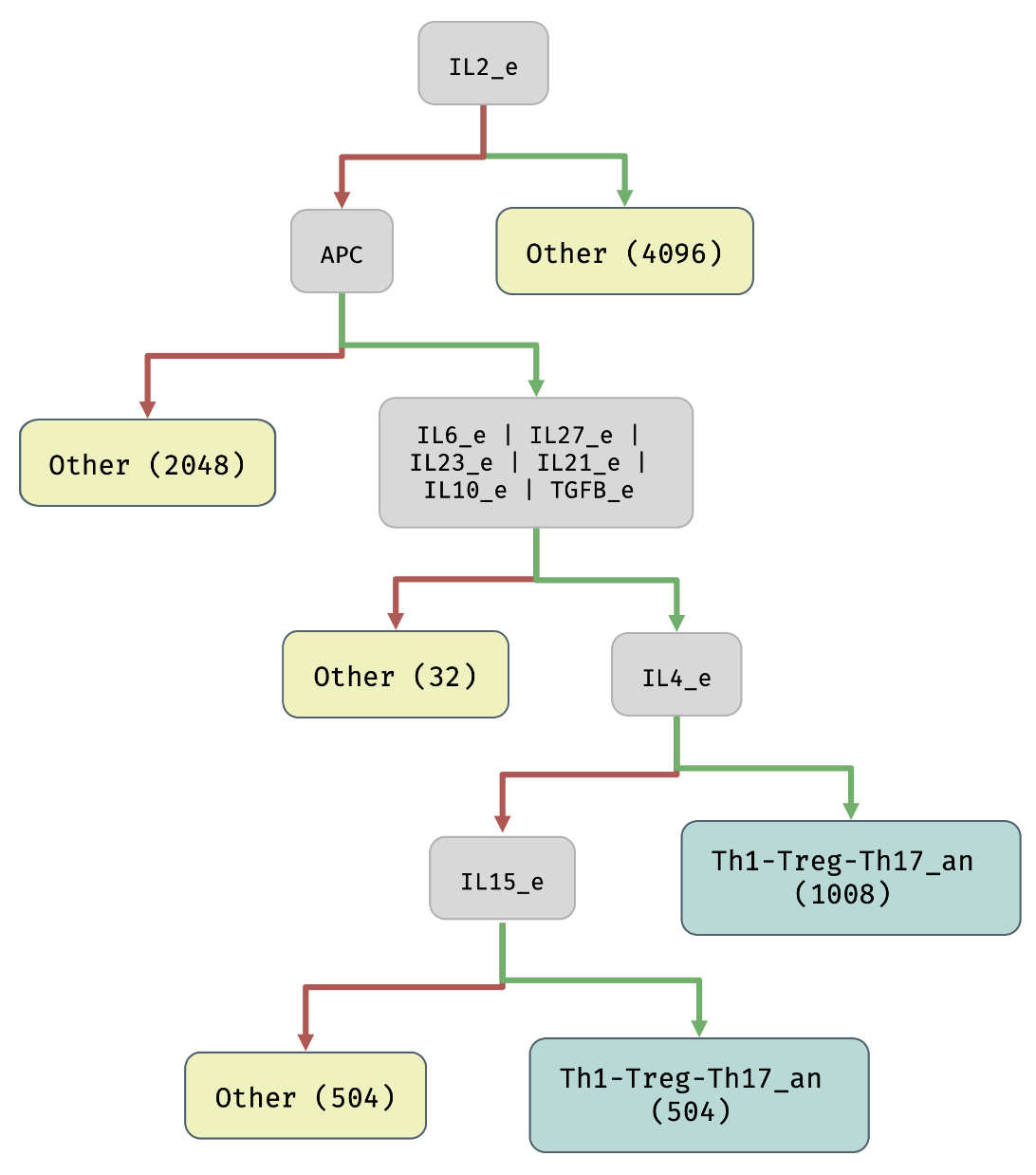}

        \vspace{3pt}
        (e) \texttt{Th1-Treg-Th17\_an}
    \end{minipage}

    \vspace{5pt}
    \caption{Decision trees visualizing the input conditions for the five most complex biologically-motivated dynamical phenotypes. Note that the condition of \texttt{Th2-Treg\_an} is equivalent to \texttt{Th1-Treg\_an}, and that of \texttt{Th2-Treg-Th17\_an} is equivalent to \texttt{Th1-Treg-Th17\_an}; consequently, these two phenotypes are not shown. Green edges mean the state 1 of the input or input combination at the start of the edge, and red edges mark the state 0. When the values of multiple inputs are given, the ``$\mid$'' symbol indicates logical OR. }
    \label{fig:input-condition-trees}
\end{figure}

The input combinations for each dynamical phenotype are derived from a disjunctive normal form identified by our symbolic workflow. Note that the maximum number of DNF clauses characterizing a single phenotype was 12, meaning each set of input conditions has a relatively compact symbolic representation. Furthermore, 9/30 phenotypes can be characterized by a single condition. In the enumeration below, necessary conditions common in all DNF clauses for that phenotype are extracted into a separate conjunctive term to minimize repetition. Finally, the most complex input conditions are visualized as decision trees in Figure~\ref{fig:input-condition-trees}.

\begin{enumerate}
    \item \texttt{Th0\_re} $\mid$ $\{ \texttt{APC} \mapsto 0, \texttt{IFNB\_e} \mapsto 0, \texttt{IFNG\_e} \mapsto 0, \texttt{IL27\_e} \mapsto 0, \texttt{IL4\_e} \mapsto 0 \}$ and:
        \begin{itemize} 
            \item $\{ \texttt{TGFB\_e} \mapsto 0\}$ or $\{ \texttt{IL10\_e} \mapsto 0, \texttt{IL21\_e} \mapsto 0, \texttt{IL6\_e} \mapsto 0 \}$.
        \end{itemize}
    
    \item \texttt{Th1\_re} $\mid$ $\{ \texttt{APC} \mapsto 0 \}$ and:
        \begin{itemize}
            \item $\{ \texttt{TGFB\_e} \mapsto 0 \}$ or $\{ \texttt{IL10\_e} \mapsto 0, \texttt{IL21\_e} \mapsto 0, \texttt{IL27\_e} \mapsto 0, \texttt{IL6\_e} \mapsto 0 \}$.
        \end{itemize}
    
    \item \texttt{Th17\_re} $\mid$ $\{ \texttt{APC} \mapsto 0, \texttt{IFNB\_e} \mapsto 0, \texttt{IFNG\_e} \mapsto 0, \texttt{IL27\_e} \mapsto 0, \texttt{IL4\_e} \mapsto 0 \}$ and:
        \begin{itemize}
            \item $\{ \texttt{TGFB\_e} \mapsto 1 \}$ or $\{ \texttt{IL6\_e} \mapsto 1 \}$ or $\{ \texttt{IL23\_e} \mapsto 1 \}$ or
            \item $\{ \texttt{IL21\_e} \mapsto 1 \}$ or $\{ \texttt{IL10\_e} \mapsto 1 \}$.
        \end{itemize}

    \item \texttt{Th1-Th17\_re} $\mid$ $\{ \texttt{APC} \mapsto 0 \}$ and:
        \begin{itemize}
            \item $\{ \texttt{TGFB\_e} \mapsto 1 \}$ or $\{ \texttt{IL6\_e} \mapsto 1 \}$ or $\{ \texttt{IL27\_e} \mapsto 1 \}$ or
            \item $\{ \texttt{IL23\_e} \mapsto 1 \}$ or $\{ \texttt{IL21\_e} \mapsto 1 \}$ or $\{ \texttt{IL10\_e} \mapsto 1 \}$.
        \end{itemize}

    \item \texttt{Th0\_ac} $\mid$ $\{  \texttt{APC} \mapsto 1, \texttt{IFNB\_e} \mapsto 0, \texttt{IFNG\_e} \mapsto 0, \texttt{IL27\_e} \mapsto 0, \texttt{IL4\_e} \mapsto 0, \texttt{TGFB\_e} \mapsto 0 \}$.

    \item \texttt{Th1\_an} $\mid$ $\{ \texttt{APC} \mapsto 1 \}$ and:
        \begin{itemize}
            \item $\{ \texttt{TGFB\_e} \mapsto 0 \}$ or
            \item $\{ \texttt{IFNG\_e} \mapsto 1, \texttt{IL10\_e} \mapsto 0, \texttt{IL21\_e} \mapsto 0, \texttt{IL27\_e} \mapsto 0, \texttt{IL2\_e} \mapsto 0, \texttt{IL6\_e} \mapsto 0 \}$ or
            \item $\{ \texttt{IFNB\_e} \mapsto 1, \texttt{IL10\_e} \mapsto 0, \texttt{IL21\_e} \mapsto 0, \texttt{IL27\_e} \mapsto 0, \texttt{IL2\_e} \mapsto 0, \texttt{IL6\_e} \mapsto 0 \}$ or 
            \item $\{ \texttt{IL10\_e} \mapsto 0, \texttt{IL15\_e} \mapsto 0, \texttt{IL21\_e} \mapsto 0, \texttt{IL27\_e} \mapsto 0, \texttt{IL2\_e} \mapsto 0, \texttt{IL4\_e}~\mapsto 0, \texttt{IL6\_e} \mapsto 0 \}$.
        \end{itemize}

    \item \texttt{Th17\_an} $\mid$ $\{ \texttt{APC} \mapsto 1, \texttt{IFNB\_e} \mapsto 0, \texttt{IFNG\_e} \mapsto 0, \texttt{IL27\_e} \mapsto 0, \texttt{IL4\_e} \mapsto 0 \}$.

    \item \texttt{Th1-Th17\_an} $\mid$ $\{ \texttt{APC} \mapsto 1 \}$ and:
        \begin{itemize}
            \item $\{ \texttt{TGFB\_e} \mapsto 1 \}$ or $\{ \texttt{IFNG\_e} \mapsto 1 \}$ or $\{ \texttt{IFNB\_e} \mapsto 1 \}$ or
            \item $\{ \texttt{IL6\_e} \mapsto 1 \}$ or $\{ \texttt{IL4\_e} \mapsto 1 \}$ or $\{ \texttt{IL2\_e} \mapsto 1 \}$ or
            \item $\{ \texttt{IL27\_e} \mapsto 1 \}$ or $\{ \texttt{IL23\_e} \mapsto 1 \}$ or $\{ \texttt{IL21\_e} \mapsto 1 \}$ or
            \item $\{ \texttt{IL15\_e} \mapsto 1 \}$ or $\{ \texttt{IL10\_e} \mapsto 1 \}$.
        \end{itemize}

    \item \texttt{Th1-Th17\_ac\_IL17} $\mid$ $\{ \texttt{APC} \mapsto 1, \texttt{IFNB\_e} \mapsto 0, \texttt{IFNG\_e} \mapsto 0, \texttt{IL15\_e} \mapsto 0, \texttt{IL27\_e} \mapsto 0, \texttt{IL2\_e} \mapsto 0, \texttt{IL4\_e} \mapsto 0 \}$.

    \item \texttt{Th1\_ac} $\mid$ $\{ \texttt{APC} \mapsto 1, \texttt{IL10\_e} \mapsto 0, \texttt{IL21\_e} \mapsto 0, \texttt{IL27\_e} \mapsto 0, \texttt{IL6\_e} \mapsto 0 \}$.

    \item \texttt{Th1-Th17\_ac\_IFNG} $\mid$ $\{ \texttt{APC} \mapsto 1, \texttt{IL10\_e} \mapsto 0, \texttt{IL21\_e} \mapsto 0, \texttt{IL27\_e} \mapsto 0, \texttt{IL6\_e} \mapsto 0, \texttt{TGFB\_e} \mapsto 1 \}$.

    \item \texttt{Th2\_re} $\mid$ $\{ \texttt{APC} \mapsto 0 \}$ and:
        \begin{itemize}
            \item $\{ \texttt{TGFB\_e} \mapsto 0 \}$ or
            \item $\{ \texttt{IL10\_e} \mapsto 0, \texttt{IL21\_e} \mapsto 0, \texttt{IL27\_e} \mapsto 0, \texttt{IL6\_e} \mapsto 0 \}$.
        \end{itemize}

    \item \texttt{Th2-Th17\_re} $\mid$ $\{ \texttt{APC} \mapsto 0 \}$ and:
        \begin{itemize}
            \item $\{ \texttt{TGFB\_e} \mapsto 1 \}$ or $\{ \texttt{IL6\_e} \mapsto 1 \}$ or $\{ \texttt{IL10\_e} \mapsto 1 \}$ or
            \item $\{ \texttt{IL27\_e} \mapsto 1 \}$ or $\{ \texttt{IL23\_e} \mapsto 1 \}$ or $\{ \texttt{IL21\_e} \mapsto 1 \}$.
        \end{itemize}

    \item \texttt{Th2\_an} $\mid$ $\{ \texttt{APC} \mapsto 1 \}$ and:
        \begin{itemize}
            \item $\{ \texttt{IL27\_e} \mapsto 1, \texttt{TGFB\_e} \mapsto 0 \}$ or
            \item $\{ \texttt{IFNG\_e} \mapsto 1, \texttt{TGFB\_e} \mapsto 0 \}$ or
            \item $\{ \texttt{IFNB\_e} \mapsto 1, \texttt{TGFB\_e} \mapsto 0 \}$ or 
            \item $\{ \texttt{IL2\_e} \mapsto 0, \texttt{IL4\_e} \mapsto 1, \texttt{TGFB\_e} \mapsto 0 \}$ or
            \item $\{ \texttt{IFNG\_e} \mapsto 1, \texttt{IL10\_e} \mapsto 0, \texttt{IL21\_e} \mapsto 0, \texttt{IL27\_e} \mapsto 0, \texttt{IL2\_e} \mapsto 0, \texttt{IL4\_e} \mapsto 1, \texttt{IL6\_e} \mapsto 0 \}$ or
            \item $\{ \texttt{IFNB\_e} \mapsto 1, \texttt{IL10\_e} \mapsto 0, \texttt{IL21\_e} \mapsto 0, \texttt{IL27\_e} \mapsto 0, \texttt{IL2\_e} \mapsto 0, \texttt{IL4\_e} \mapsto 1, \texttt{IL6\_e} \mapsto 0 \}$.
        \end{itemize}

    \item \texttt{Th2-Th17\_an} $\mid$ $\{ \texttt{APC} \mapsto 1 \}$ and:
        \begin{itemize}
            \item $\{ \texttt{IL27\_e} \mapsto 1 \}$ or $\{ \texttt{IFNG\_e} \mapsto 1 \}$ or $\{ \texttt{IFNB\_e} \mapsto 1 \}$ or
            \item $\{ \texttt{IL2\_e} \mapsto 0, \texttt{IL4\_e} \mapsto 1, \texttt{IL6\_e} \mapsto 1 \}$ or
            \item $\{ \texttt{IL23\_e} \mapsto 1, \texttt{IL2\_e} \mapsto 0, \texttt{IL4\_e} \mapsto 1 \}$ or
            \item $\{ \texttt{IL21\_e} \mapsto 1, \texttt{IL2\_e} \mapsto 0, \texttt{IL4\_e} \mapsto 1 \}$ or
            \item $\{ \texttt{IL10\_e} \mapsto 1, \texttt{IL2\_e} \mapsto 0, \texttt{IL4\_e} \mapsto 1 \}$.
        \end{itemize}

    \item \texttt{Th2\_ac} $\mid$ $\{ \texttt{APC} \mapsto 1, \texttt{IFNB\_e} \mapsto 0, \texttt{IFNG\_e} \mapsto 0, \texttt{IL27\_e} \mapsto 0, \texttt{TGFB\_e} \mapsto 0 \}$.

    \item \texttt{Th2-Th17\_ac\_IL4} $\mid$ $\{ \texttt{APC} \mapsto 1, \texttt{IFNB\_e} \mapsto 0, \texttt{IFNG\_e} \mapsto 0, \texttt{IL27\_e} \mapsto 0 \}$.

    \item \texttt{Treg\_an} $\mid$ $\{ \texttt{APC} \mapsto 1, \texttt{IFNB\_e} \mapsto 0, \texttt{IFNG\_e} \mapsto 0, \texttt{IL15\_e} \mapsto 1, \texttt{IL27\_e} \mapsto 0, \texttt{IL2\_e} \mapsto 0, \texttt{IL4\_e} \mapsto 0\}$ and:
        \begin{itemize}
            \item $\{ \texttt{TGFB\_e} \mapsto 0 \}$ or
            \item $\{ \texttt{IL10\_e} \mapsto 0, \texttt{IL21\_e} \mapsto 0, \texttt{IL6\_e} \mapsto 0 \}$.
        \end{itemize}

    \item \texttt{Treg\_ac} $\mid$ $\{ \texttt{APC} \mapsto 1, \texttt{IFNB\_e} \mapsto 0, \texttt{IFNG\_e} \mapsto 0, \texttt{IL10\_e} \mapsto 0, \texttt{IL21\_e} \mapsto 0, \texttt{IL27\_e} \mapsto 0, \texttt{IL4\_e} \mapsto 0, \texttt{IL6\_e} \mapsto 0 \}$ and:
        \begin{itemize}
            \item $\{ \texttt{IL2\_e} \mapsto 1 \}$ or $\{ \texttt{IL15\_e} \mapsto 1 \}$.
        \end{itemize}

    \item \texttt{Th1-Treg\_an} $\mid$ $\{ \texttt{APC} \mapsto 1, \texttt{IL2\_e} \mapsto 0 \}$ and:
        \begin{itemize}
            \item $\{ \texttt{IL4\_e} \mapsto 1, \texttt{TGFB\_e} \mapsto 0 \}$ or $\{ \texttt{IL15\_e} \mapsto 1, \texttt{TGFB\_e} \mapsto 0 \}$ or
            \item $\{ \texttt{IL10\_e} \mapsto 0, \texttt{IL21\_e} \mapsto 0, \texttt{IL27\_e} \mapsto 0, \texttt{IL4\_e} \mapsto 1, \texttt{IL6\_e} \mapsto 0 \}$ or
            \item $\{ \texttt{IL10\_e} \mapsto 0, \texttt{IL15\_e} \mapsto 1, \texttt{IL21\_e} \mapsto 0, \texttt{IL27\_e} \mapsto 0, \texttt{IL4\_e} \mapsto 0, \texttt{IL6\_e} \mapsto 0 \}$.
        \end{itemize}

    \item \texttt{Th1-Treg\_ac\_TGFB} $\mid$ $\{ \texttt{APC} \mapsto 1, \texttt{IL10\_e} \mapsto 0, \texttt{IL21\_e} \mapsto 0, \texttt{IL27\_e} \mapsto 0, \texttt{IL6\_e} \mapsto 0 \}$ and:
        \begin{itemize}
            \item $\{ \texttt{IL4\_e} \mapsto 1 \}$ or $\{ \texttt{IL2\_e} \mapsto 1 \}$ or $\{ \texttt{IL15\_e} \mapsto 1 \}$.
        \end{itemize}

    \item \texttt{Treg-Th17\_an} $\mid$ $\{ \texttt{APC} \mapsto 1, \texttt{IFNB\_e} \mapsto 0, \texttt{IFNG\_e} \mapsto 0, \texttt{IL15\_e} \mapsto 1, \texttt{IL27\_e} \mapsto 0, \texttt{IL2\_e} \mapsto 0, \texttt{IL4\_e} \mapsto 0 \}$ and:
        \begin{itemize}
            \item $\{ \texttt{TGFB\_e} \mapsto 1 \}$ or $\{ \texttt{IL6\_e} \mapsto 1 \}$ or $\{ \texttt{IL23\_e} \mapsto 1 \}$ or
            \item $\{ \texttt{IL21\_e} \mapsto 1\}$ or $\{ \texttt{IL10\_e} \mapsto 1 \}$.
        \end{itemize}

    \item \texttt{Treg-Th17\_ac\_TGFB} $\mid$ $\{ \texttt{APC} \mapsto 1, \texttt{IFNB\_e} \mapsto 0, \texttt{IFNG\_e} \mapsto 0, \texttt{IL27\_e} \mapsto 0, \texttt{IL4\_e} \mapsto 0 \}$ and:
        \begin{itemize}
            \item $\{ \texttt{IL2\_e} \mapsto 1 \}$ or $\{ \texttt{IL15\_e} \mapsto 1 \}$.
        \end{itemize}

    \item \texttt{Th1-Treg-Th17\_an} $\mid$ $\{ \texttt{APC} \mapsto 1, \texttt{IL2\_e} \mapsto 0 \}$ and:
        \begin{itemize}
            \item $\{ \texttt{IL4\_e} \mapsto 1, \texttt{TGFB\_e} \mapsto 1 \}$ or
            \item $\{ \texttt{IL4\_e} \mapsto 1, \texttt{IL6\_e} \mapsto 1 \}$ or
            \item $\{ \texttt{IL15\_e} \mapsto 1, \texttt{TGFB\_e} \mapsto 1 \}$ or
            \item $\{ \texttt{IL15\_e} \mapsto 1, \texttt{IL6\_e} \mapsto 1 \}$ or
            \item $\{ \texttt{IL27\_e} \mapsto 1, \texttt{IL4\_e} \mapsto 1 \}$ or
            \item $\{ \texttt{IL15\_e} \mapsto 1, \texttt{IL27\_e} \mapsto 1, \texttt{IL4\_e} \mapsto 0 \}$ or
            \item $\{ \texttt{IL23\_e} \mapsto 1, \texttt{IL4\_e} \mapsto 1 \}$ or
            \item $\{ \texttt{IL15\_e} \mapsto 1, \texttt{IL23\_e} \mapsto 1, \texttt{IL4\_e} \mapsto 0 \}$ or
            \item $\{ \texttt{IL21\_e} \mapsto 1, \texttt{IL4\_e} \mapsto 1 \}$ or
            \item $\{ \texttt{IL15\_e} \mapsto 1, \texttt{IL21\_e} \mapsto 1, \texttt{IL4\_e} \mapsto 0 \}$ or
            \item $\{ \texttt{IL10\_e} \mapsto 1, \texttt{IL4\_e} \mapsto 1 \}$ or
            \item $\{ \texttt{IL10\_e} \mapsto 1, \texttt{IL15\_e} \mapsto 1 \}$.
        \end{itemize}

    \item \texttt{Th1-Treg-Th17\_ac\_TGFB} $\mid$ $\{ \texttt{APC} \mapsto 1 \}$ and:
        \begin{itemize}
            \item $\{ \texttt{IL2\_e} \mapsto 1 \}$ or $\{ \texttt{IL4\_e} \mapsto 1 \}$ or $\{ \texttt{IL15\_e} \mapsto 1 \}$.
        \end{itemize}

    \item \texttt{Th2-Treg\_an} $\mid$ $\{ \texttt{APC} \mapsto 1, \texttt{IL2\_e} \mapsto 0 \}$ and:
        \begin{itemize}
            \item $\{ \texttt{IL4\_e} \mapsto 1, \texttt{TGFB\_e} \mapsto 0 \}$ or
            \item $\{ \texttt{IL15\_e} \mapsto 1, \texttt{TGFB\_e} \mapsto 0 \}$ or
            \item $\{ \texttt{IL10\_e} \mapsto 0, \texttt{IL21\_e} \mapsto 0, \texttt{IL27\_e} \mapsto 0, \texttt{IL4\_e} \mapsto 1, \texttt{IL6\_e} \mapsto 0  \}$ or
            \item $\{ \texttt{IL10\_e} \mapsto 0, \texttt{IL15\_e} \mapsto 1, \texttt{IL21\_e} \mapsto 0, \texttt{IL27\_e} \mapsto 0, \texttt{IL4\_e} \mapsto 0, \texttt{IL6\_e} \mapsto 0 \}$.
        \end{itemize}

    \item \texttt{Th2-Treg-Th17\_an} $\mid$ $\{ \texttt{APC} \mapsto 1, \texttt{IL2\_e} \mapsto 0 \}$ and:
        \begin{itemize}
            \item $\{ \texttt{IL4\_e} \mapsto 1, \texttt{TGFB\_e} \mapsto 1 \}$ or
            \item $\{ \texttt{IL4\_e} \mapsto 1, \texttt{IL6\_e} \mapsto 1 \}$ or
            \item $\{ \texttt{IL15\_e} \mapsto 1, \texttt{TGFB\_e} \mapsto 1 \}$ or
            \item $\{ \texttt{IL15\_e} \mapsto 1, \texttt{IL6\_e} \mapsto 1 \}$ or
            \item $\{ \texttt{IL27\_e} \mapsto 1, \texttt{IL4\_e} \mapsto 1 \}$ or
            \item $\{ \texttt{IL15\_e} \mapsto 1, \texttt{IL27\_e} \mapsto 1, \texttt{IL4\_e} \mapsto 0 \}$ or
            \item $\{ \texttt{IL23\_e} \mapsto 1, \texttt{IL4\_e} \mapsto 1 \}$ or
            \item $\{ \texttt{IL15\_e} \mapsto 1, \texttt{IL23\_e} \mapsto 1, \texttt{IL4\_e} \mapsto 0 \}$ or
            \item $\{ \texttt{IL21\_e} \mapsto 1, \texttt{IL4\_e} \mapsto 1 \}$ or
            \item $\{ \texttt{IL15\_e} \mapsto 1, \texttt{IL21\_e} \mapsto 1, \texttt{IL4\_e} \mapsto 0 \}$ or
            \item $\{ \texttt{IL10\_e} \mapsto 1, \texttt{IL4\_e} \mapsto 1 \}$ or
            \item $\{ \texttt{IL10\_e} \mapsto 1, \texttt{IL15\_e} \mapsto 1 \}$.
        \end{itemize}

    \item \texttt{Th2-Treg-Th17\_ac\_TGFB} $\mid$ $\{ \texttt{APC} \mapsto 1 \}$ and:
        \begin{itemize}
            \item $\{ \texttt{IL2\_e} \mapsto 1 \}$ or $\{ \texttt{IL4\_e} \mapsto 1 \}$ or $\{ \texttt{IL15\_e} \mapsto 1 \}$.
        \end{itemize}

    \item \texttt{Treg\_ac\_o} $\mid$ $\{ \texttt{APC} \mapsto 1, \texttt{IFNB\_e} \mapsto 0, \texttt{IFNG\_e} \mapsto 0, \texttt{IL10\_e} \mapsto 0, \texttt{IL15\_e} \mapsto 0, \texttt{IL21\_e} \mapsto 0, \texttt{IL27\_e} \mapsto 0, \texttt{IL2\_e} \mapsto 0, \texttt{IL4\_e} \mapsto 0, \texttt{IL6\_e} \mapsto 0, \texttt{TGFB\_e} \mapsto 1 \}$.

    \item \texttt{Treg-Th17\_ac\_TGFB\_o} $\mid$ $\{ \texttt{APC} \mapsto 1, \texttt{IFNB\_e} \mapsto 0, \texttt{IFNG\_e} \mapsto 0, \texttt{IL10\_e} \mapsto 0, \texttt{IL15\_e} \mapsto 0, \texttt{IL21\_e} \mapsto 0, \texttt{IL27\_e} \mapsto 0, \texttt{IL2\_e} \mapsto 0, \texttt{IL4\_e} \mapsto 0, \texttt{IL6\_e} \mapsto 0, \texttt{TGFB\_e} \mapsto 1
 \}$.
        
\end{enumerate}

\section{Supplementary figures and tables}

\begin{figure}
    \centering
    \includegraphics[width=0.45\linewidth]{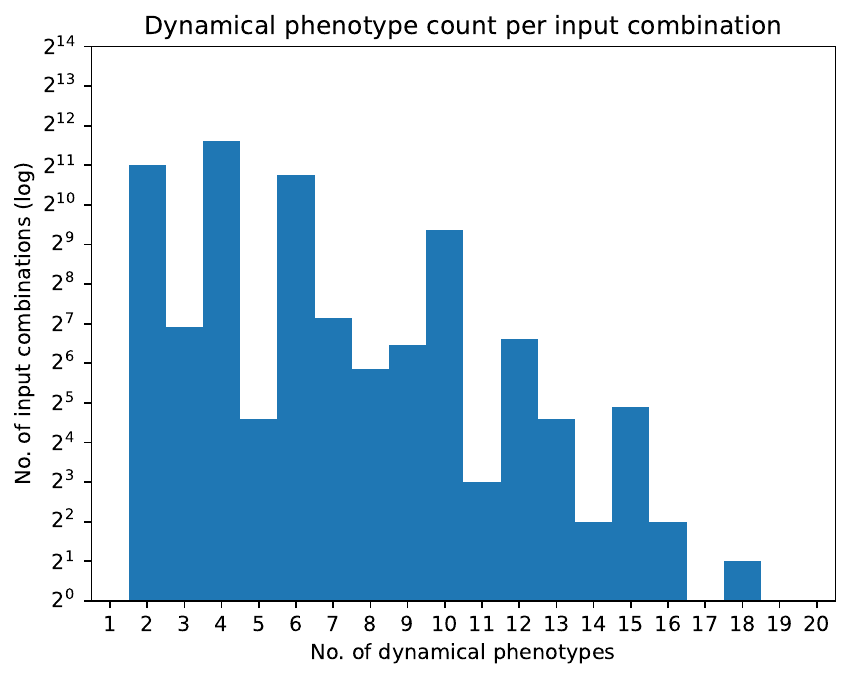}
    \includegraphics[width=0.45\linewidth]{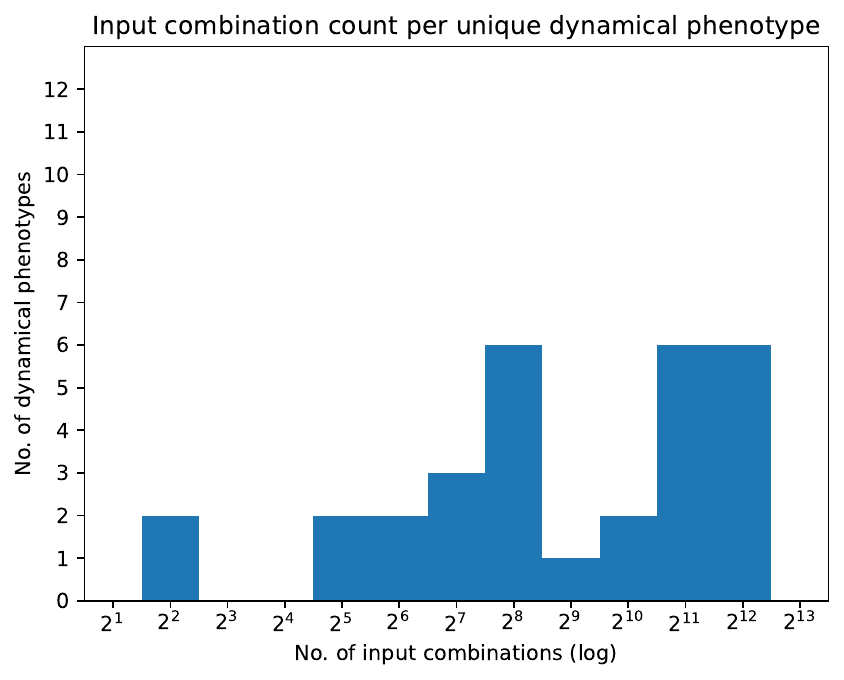}

    \includegraphics[width=0.45\linewidth]{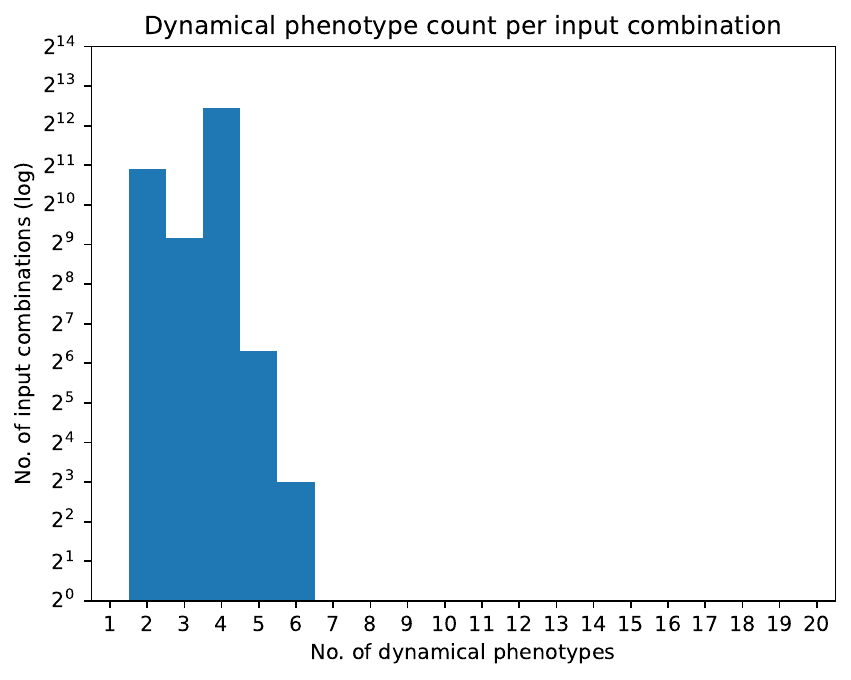}
    \includegraphics[width=0.45\linewidth]{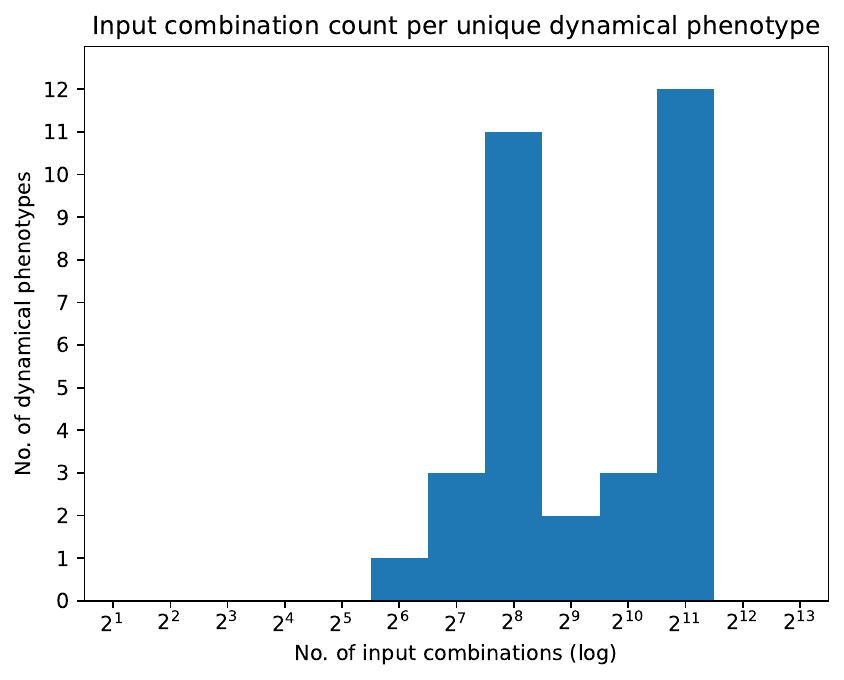}
    
    \caption{Histograms of the dynamical phenotype versus input combination counts for the biologically-motivated (top) and LDOI-based (bottom) PDNs. Note that in both graphs, the number of input valuations is presented on logarithmic scales (y-axis on the left, x-axis on the right).}
    \label{fig:phenotype-count-histograms}
\end{figure}

\begin{table}[]

    \begin{tabular}{L{110pt} R{50pt} R{50pt}}
        \toprule
        \textbf{Phenotype Name} & \textbf{No. of subspaces} & \textbf{Avg. Entropy} \\
        \midrule
        Th0\_ac & 1 & 0 \\
        Th0\_re & 2 & 0.128761 \\
        Th1-Th17\_ac\_IFNG & 1 & 0 \\
        Th1-Th17\_ac\_IL17 & 1 & 0 \\
        Th1-Th17\_an & 29 & 0.430095 \\
        Th1-Th17\_re & 6 & 0.142286 \\
        Th1-Treg-Th17\_ac\_TGFB & 3 & 0.153713 \\
        Th1-Treg-Th17\_an & 12 & 0.245295 \\
        Th1-Treg\_ac\_TGFB & 3 & 0.153713 \\
        Th1-Treg\_an & 4 & 0.257522 \\
        Th17\_an & 1 & 0 \\
        Th17\_re & 5 & 0.111548 \\
        Th1\_ac & 1 & 0 \\
        Th1\_an & 9 & 0.451735 \\
        Th1\_re & 2 & 0.154513 \\
        Th2-Th17\_ac\_IL4 & 1 & 0 \\
        Th2-Th17\_an & 17 & 0.387609 \\
        Th2-Th17\_re & 6 & 0.142286 \\
        Th2-Treg-Th17\_ac\_TGFB & 3 & 0.141889 \\
        Th2-Treg-Th17\_an & 12 & 0.232419 \\
        Th2-Treg\_an & 4 & 0.244646 \\
        Th2\_ac & 1 & 0 \\
        Th2\_an & 6 & 0.454081 \\
        Th2\_re & 2 & 0.154513 \\
        Treg-Th17\_ac\_TGFB & 2 & 0.090133 \\
        Treg-Th17\_ac\_TGFB\_o & 1 & 0 \\
        Treg-Th17\_an & 5 & 0.111548 \\
        Treg\_ac & 2 & 0.090133 \\
        Treg\_ac\_o & 1 & 0 \\
        Treg\_an & 2 & 0.128761 \\
        \midrule
        & Average & 0.146906 \\
        \bottomrule
    \end{tabular}

    \vspace{8pt}
    \caption{The number of dynamical phenotype subspaces contained in each biologically-motivated dynamical phenotype (see Table~\ref{Bio_phenotypes}) and the average normalized marginal entropy of the non-PDN variables within these subspaces. Entropy zero is a perfectly homogeneous set, entropy one corresponds to uniform distribution of $\{0,1,\any\}$ values.}
    \label{tab:phenotype-entropy-literature}
\end{table}

\begin{table}[]
    \begin{tabular}{L{80pt} R{50pt} R{50pt}}
        \toprule
        \textbf{Phenotype Name} & \textbf{No. of subspaces} & \textbf{Avg. Entropy} \\
        \midrule
        00000 & 1 & 0 \\
        00001 & 1 & 0 \\
        00010 & 3 & 0.076522 \\
        00011 & 3 & 0.076522 \\
        00100 & 1 & 0 \\
        00101 & 1 & 0 \\
        00110 & 3 & 0.076522 \\
        00111 & 3 & 0.076522 \\
        01000 & 1 & 0 \\
        01001 & 3 & 0.185839 \\
        01010 & 3 & 0.076522 \\
        01011 & 3 & 0.06559 \\
        01100 & 7 & 0.15605 \\
        01101 & 3 & 0.240497 \\
        01110 & 6 & 0.136044 \\
        01111 & 4 & 0.298163 \\
        10000 & 1 & 0 \\
        10001 & 1 & 0 \\
        10010 & 3 & 0.076522 \\
        10011 & 3 & 0.076522 \\
        10100 & 1 & 0 \\
        10101 & 1 & 0 \\
        10110 & 3 & 0.076522 \\
        10111 & 3 & 0.076522 \\
        11000 & 1 & 0 \\
        11001 & 3 & 0.109317 \\
        11010 & 3 & 0.076522 \\
        11011 & 3 & 0.06559 \\
        11100 & 1 & 0 \\
        11101 & 2 & 0.130948 \\
        11110 & 3 & 0.076522 \\
        11111 & 4 & 0.278848 \\
        \midrule
        & Average & 0.078394 \\
        \bottomrule
    \end{tabular}

    \vspace{8pt}
    \caption{The number of dynamical phenotype subspaces contained in each LDOI-based dynamical phenotype and the average normalized marginal entropy of the non-PDN variables within these subspaces. Entropy zero is a perfectly homogeneous set, entropy one corresponds to uniform distribution of $\{0,1,\any\}$ values. The LDOI phenotypes are denoted by a binary string indicating the state of IL4R\_b1, NFAT, STAT1, TBET, and proliferation, respectively.}
    \label{tab:phenotype-entropy-ldoi}
\end{table}

\begin{table}[]
\begingroup
    \setlength{\tabcolsep}{10pt} 
    \renewcommand{\arraystretch}{1.5} 
    \small
    \centering
    \begin{tabular}{L{1.4cm}L{1cm}L{3.7cm}L{2.1cm}L{1.1cm}}
IL4R\_b1, STAT1 & Notation & Nodes fixed by LDOI & Additional nodes fixed in all trap spaces & Active STAT nodes\\\hline\hline
1, 0 & A & IL12RB2=0, IRF1=0, IL12RB1=0, IL12R=0, IL17=0, STAT4=0, STAT5\_b1=1, STAT6=1 & IFNG=0, IL2=0 & STAT5, STAT6\\\hline
1, 1 & B & IL12RB2=0, IRF=1, IL12RB1=1, IL12R=0, IL17=0, STAT4=0, STAT5\_b1=1, STAT6=1 & IL4=0, IL2=0 & STAT1, STAT5, STAT6\\\hline
0, 1 & C & IL4R\_b2=0, IL12RB2=1,IRF1=1, IL12RB1=1, IL17=0, STAT6=0 & IL4=0 & STAT1\\\hline
0, 0 & D & IL4R\_b2=0, IL12RB2=1, IRF1=0, IL12RB1=0, IL12R=0, STAT4=0, STAT6=0 & IFNG=0, IL4=0 & \\\hline
    \end{tabular}
\endgroup
\vspace{8pt}
    \caption{Nodes fixed in dynamical phenotype groups that have the same value of IL4R\_b1 and STAT1 (e.g., in the 8 dynamical phenotypes labeled A in Figure~\ref{fig:LDOIvsbiophenotype_matrix}). The fourth column indicates that even more nodes have a fixed value than those due to the canalizing influences of these two PDNs. The last column highlights the STAT proteins that are activated in each of these dynamical phenotype group.}
    \label{LDOI_families}
\end{table}

\begin{figure}
    \centering
    \begin{minipage}{0.8\linewidth}
        \includegraphics[width=1.0\linewidth]{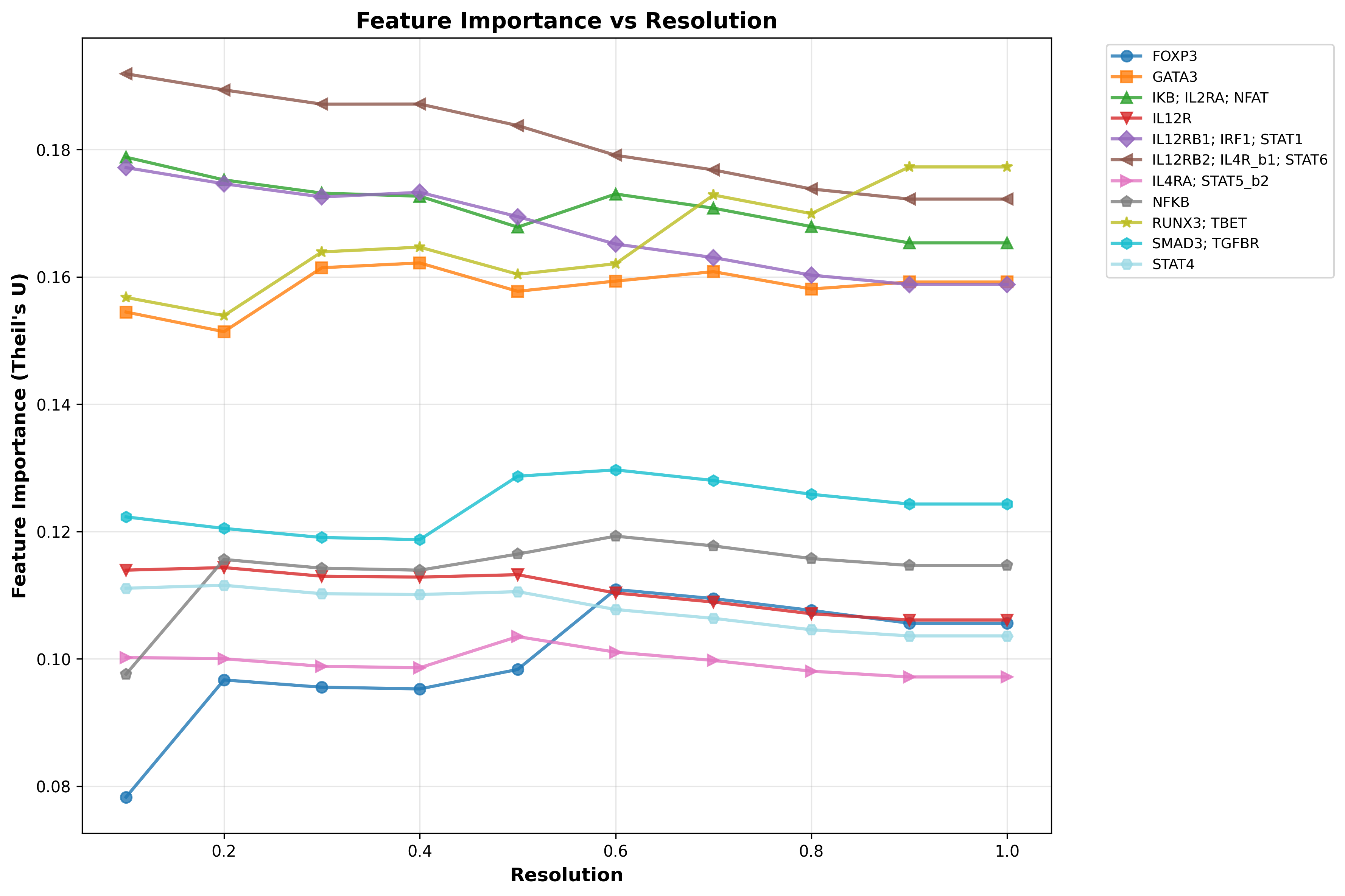}
    \end{minipage}    
    \begin{minipage}{0.8\linewidth}
        \includegraphics[width=1.0\linewidth]{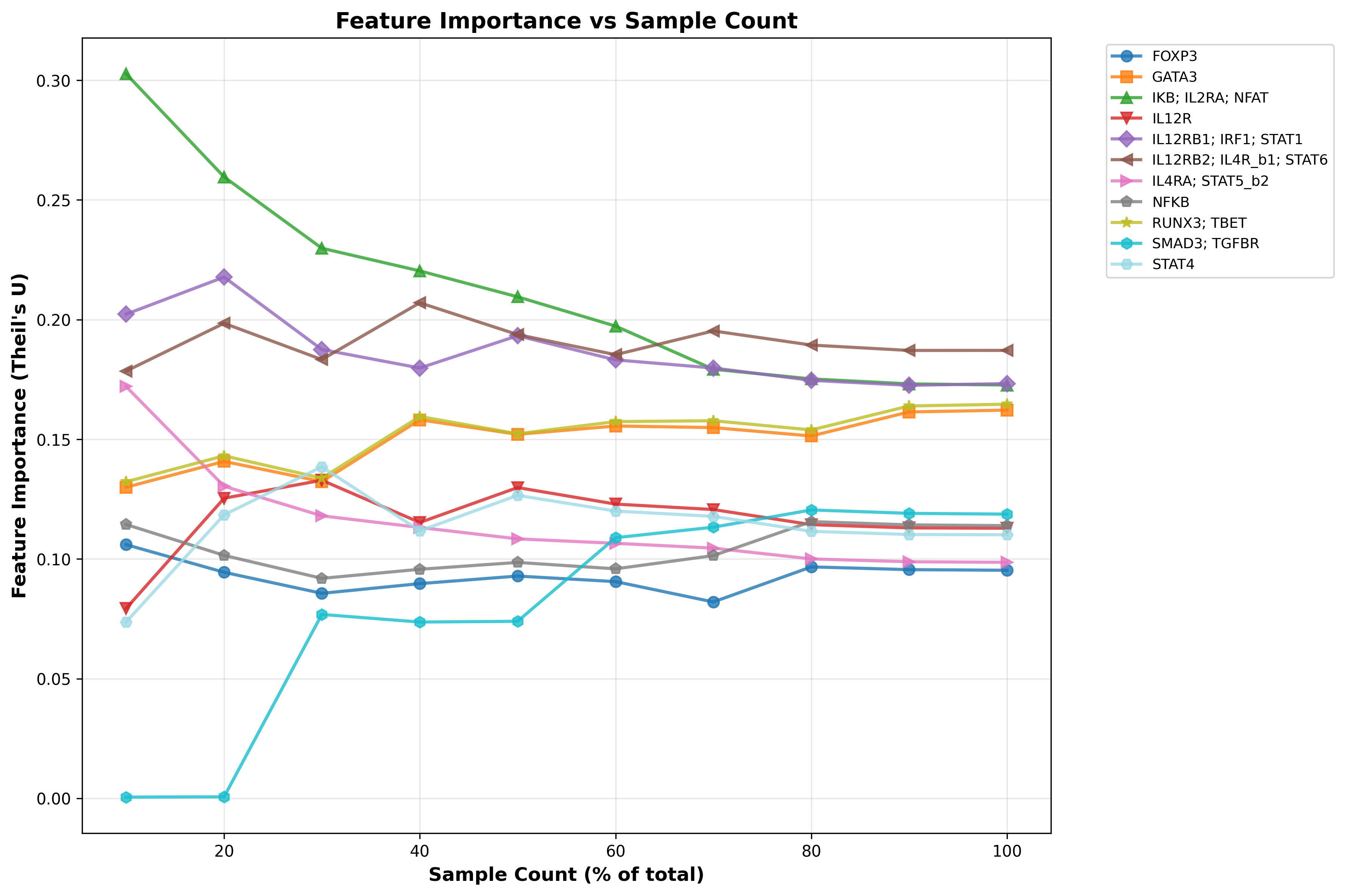}
    \end{minipage}    
    \caption{The relationship between feature importance of model variables vs. resolution (top; considering all attractors) and attractor sample size (bottom; fixed resolution $0.4$). Nodes with equivalent importance across all data points are joined into a single series (indicated by ``;'' in figure legend).}
    \label{fig:feature-importance-variations}
\end{figure}

\begin{figure}
    \centering    
    \includegraphics[width=1.0\linewidth, trim={1.0cm 1.8cm 1.0cm 1.8cm}, clip]{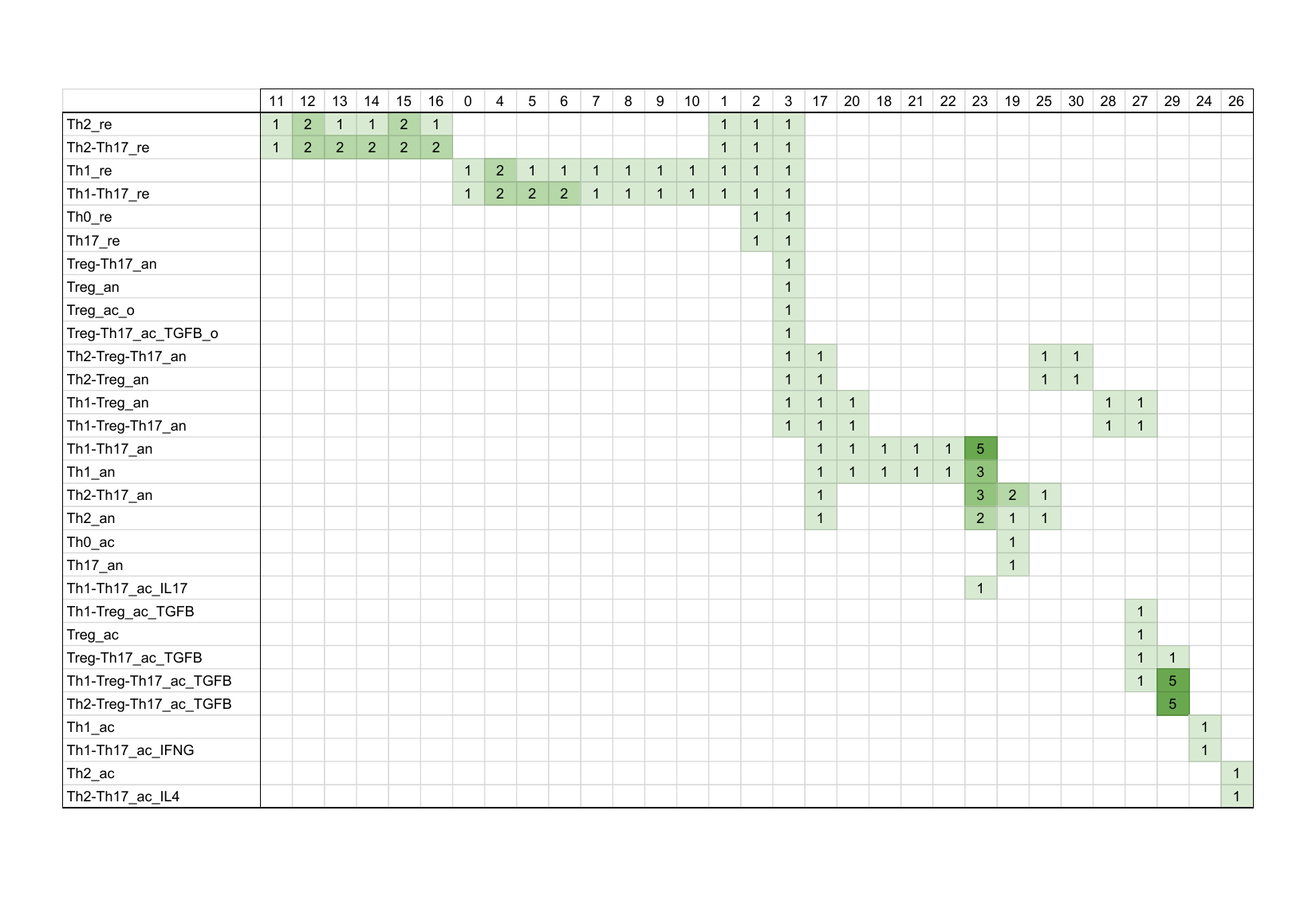}
    \caption{Mapping between the biologically-motivated phenotypes (rows; see also Table~\ref{Bio_phenotypes}) and the 31 attractor clusters (columns). The numbers and color intensity indicate the attractor count in thousands ($1 \equiv [1, 1000]$, $2 \equiv [1000, 1999]$, etc.).}
    \label{fig:clustervsbiophenotype_matrix}
\end{figure}

\end{appendices}

\end{document}